\documentclass[aps,prb,superscriptaddress,twocolumn,longbibliography,preprintnumbers,amsmath,amssymb]{revtex4-1}
\usepackage{color}
\usepackage{array}
\usepackage{amsmath}
\usepackage{amssymb}
\usepackage[pdftex]{graphicx}
\usepackage{epstopdf}
\usepackage{bbm}
\usepackage{hyperref}
\usepackage{float}

\newcommand{\gG}{s}
\newcommand{\bra}[1]{\langle #1|}
\newcommand{\ket}[1]{|#1\rangle}

\begin{document}
\title{Generalized modular transformations in 3+1D topologically ordered phases and triple linking invariant of loop braiding}
\author{Shenghan Jiang, Andrej Mesaros and Ying Ran}
\affiliation{Department of Physics, Boston College, Chestnut Hill, MA 02467}

\begin{abstract}
  In topologically ordered quantum states of matter in 2+1D (space-time dimensions), the braiding statistics of anyonic quasiparticle excitations is a fundamental characterizing property which is directly related to global transformations of the ground-state wavefunctions on a torus (the modular transformations). On the other hand, there are theoretical descriptions of various topologically ordered states in 3+1D, which exhibit both point-like and loop-like excitations, but systematic understanding of the fundamental physical distinctions between phases, and how these distinctions are connected to quantum statistics of excitations, is still lacking. One main result of this work is that the three-dimensional generalization of modular transformations, when applied to topologically ordered ground states, is directly related to a certain braiding process of loop-like excitations. This specific braiding surprisingly involves \textit{three} loops simultaneously, and can distinguish different topologically ordered states.
% To show this, we consider a basis of minimally entangled states and the algebra of membrane operators which describes the loop braiding.
Our second main result is the identification of the three-loop braiding as a process in which the worldsheets of the three loops have a non-trivial \textit{triple linking number}, which is a topological invariant characterizing closed two-dimensional surfaces in four dimensions. In this work we consider realizations of topological order in 3+1D using cohomological gauge theory in which the loops have Abelian statistics, and explicitly demonstrate our results on examples with $Z_2\times Z_2$ topological order.
\end{abstract}

\date{\today}

\maketitle
\tableofcontents

\section{Introduction}

\begin{figure}
\includegraphics[width=0.35\textwidth]{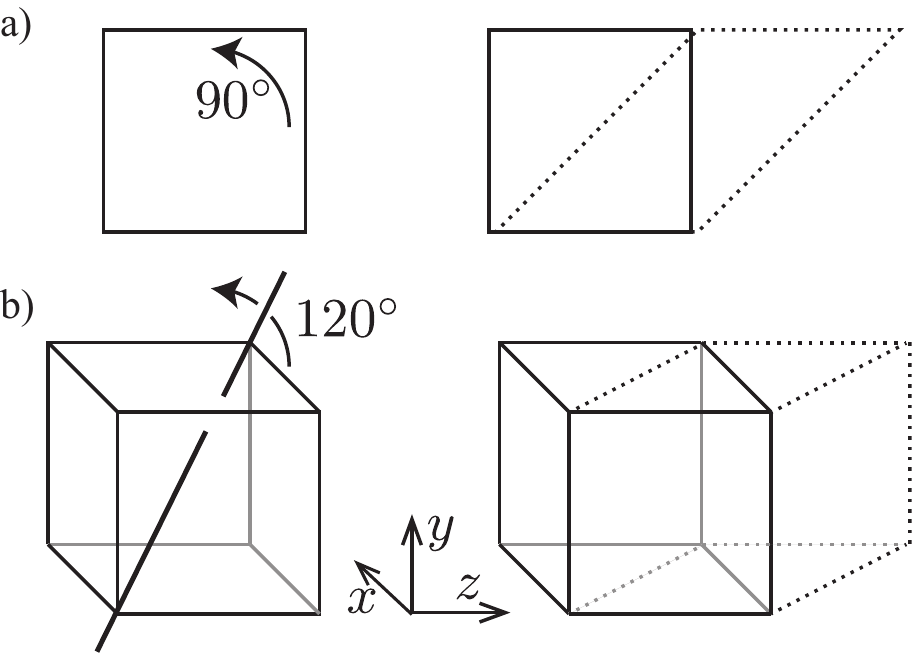}
\caption{$S$ (left) and $T$ (right) transformations on the (a) Two-torus and (b) Three-torus, which are defined by periodic boundary conditions.}
  \label{fig:2dST}
\end{figure}

Topologically ordered quantum phases of matter in 2+1D have been intriguing since their discovery decades ago (see \onlinecite{Wen:1995p6287} and references therein), due to exotic properties such as fractionalized quasiparticles with anyonic quantum braiding statistics.\cite{Wen:1990p7458,Wen:1990p5870} Early on it was realized that in such phases the topological degeneracy of the ground state on the torus corresponds to the number of types of particle excitations (superselection sectors).\cite{Wen:1992p6753} Furthermore, it was shown that the matrix of Berry's phases experienced by the ground states under the modular transformations of the torus, the $S$ and $T$ transformations (Fig.~\ref{fig:2dST}a), are directly related to the quantum statistics of the quasiparticles.\cite{Wen:1990p7458} In fact, to date the most fundamental conjecture remains that the matrices of $S,T$ contain complete information about a topological order.\cite{Wen:1990p7458} Therefore one can view the modular $S,T$ matrices as the ``non-local order parameters'' in a topologically ordered phase.\cite{Bais:2014p7544}

However, in three spatial dimensions some fundamental questions are yet completely unresolved: Is there a physical way to characterize different topological orders in 3+1D? Can braiding of excitations help us in the characterization? Clearly the problem is much more complex, since in 3d there are generically both point-like and loop-like excitations, and their geometric interplay is rich. If some type of braiding can help us characterize the topological order in 3+1D, what is the topological property of that braiding process that is relevant?

Motivated by the fundamental role of modular transformations of the torus in 2+1D systems, our approach to these questions is based on considering the analogous transformations on the three-torus (e.g., a cube with periodic boundary conditions). The modular transformations $S,T$ on the torus generate the group $SL(2,Z)$, which represents the different classes of continuous transformations on the torus.\footnote{More precisely, $SL(2,Z)$ is the mapping class group of the two-torus, i.e., the group of isotopy-classes of automorphisms of the torus. The mapping class group is formed by Dehn twists of the torus.} In 3+1D quantum states, the analogue is the three-torus, which also has just two associated transformations $S,T$, generators of $SL(3,Z)$ group,\cite{Trott:1962p8004,Moradi:2014p7978} namely a $120^\circ$ rotation through a diagonal of the periodic cube and a shear, respectively (Fig.~\ref{fig:2dST}b). Very recently it has been conjectured that exactly these kinds of transformations can be used to characterize topological order in any dimension.\cite{Moradi:2014p7978}

One way to study topologically ordered states is using the exactly solvable models of discrete gauge theories introduced by Dijkgraaf and Witten (DW).\cite{Dijkgraaf:1990p7194,Propitius:1993CS_ccy} Although these theories in 2+1D do not provide an exhaustive classification of all possible topological orders,\footnote{Discrete gauge theories can only describe non-chiral states having quasiparticles with integer quantum dimension. Also, some distinct phases can differ by a physically irrelevant relabeling of quasiparticles.} they describe a physically interesting set of states. Most importantly for this work, such cohomological gauge theories with gauge group $G$ are naturally defined in any spatial dimension, allowing us to study 3+1D topological orders. They also host both point-like and loop-like excitations, namely gauge charges and flux-loops, respectively. For simplicity, we restrict to the case of Abelian groups $G$, and then additionally to cases where loops have only Abelian braiding.

\label{sec:braid_S}
\begin{figure}
  \includegraphics[width=0.48\textwidth]{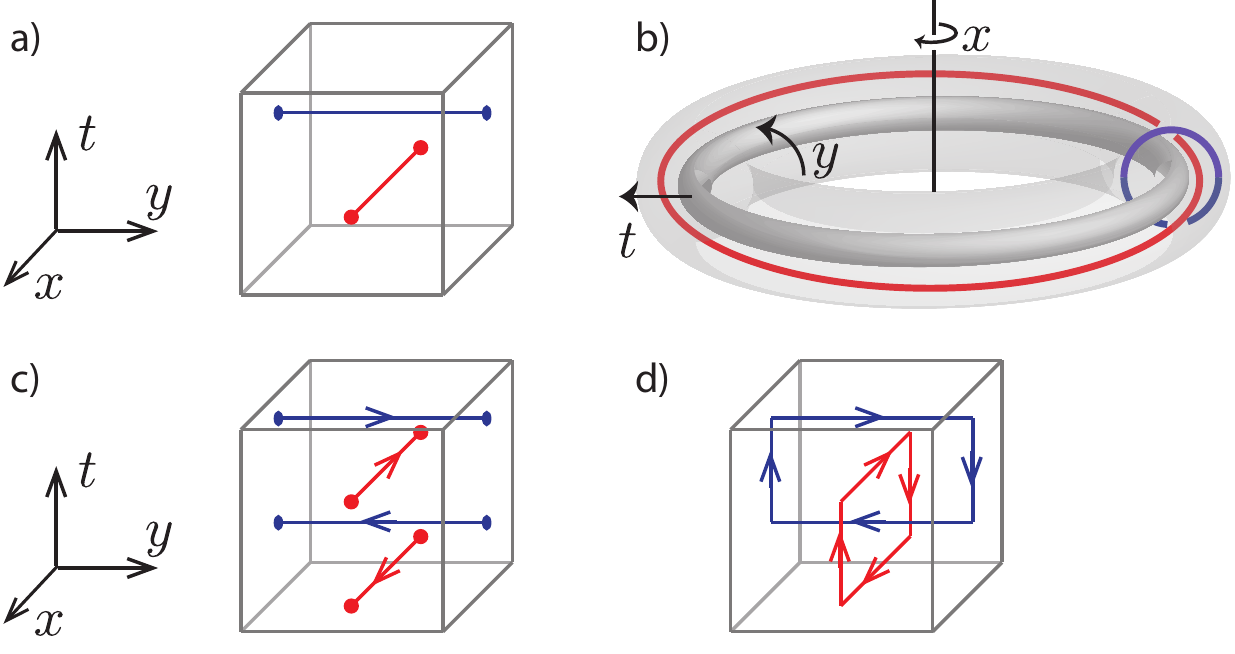}
  \caption{An $\mathcal{S}$ matrix element and braiding in 2+1D. (a) Matrix element equals an overlap of two MES, shown as a time sequence. Any MES is defined by action of particle tunneling operator along $x$ on the appropriate reference state defined by $x$ direction. The MES at later time (blue) has been acted on by $S$ modular transformation, so the tunneling is along $y$, and the final reference state is defined by $y$ direction. (b) Embedding the space-time process of (a) in three dimensions shows that the two worldlines are linked. Time grows in the radial direction as shown, so the space-time of (a) spans the volume of a toroidal slab. (c) Alternatively, the matrix element equals a product of tunneling operators, also presented as a time sequence. Arrows mark the action of tunneling operator and its inverse. Due to taking the expectation value of this operator product, the initial and final state are the same, in contrast to panel (a). (d) Connecting the worldlines from (c) one obtains two worldlines that are linked.}
  \label{fig:2dbraid}
\end{figure}

Our approach to the 3+1D problem is based on generalizing some aspects of the 2+1D case, which we review here. It is well known that topological operators, which describe tunneling of quasiparticle across the periodic 2d system, obey a non-trivial algebra. A certain product of these operators gives the identity operator times a complex number which equals an $\mathcal{S}$ matrix element, and also equals the quasiparticle statistics.\cite{Wen:1990p5870,Oshikawa:2006p5869} To generalize to three dimensions, we need the topological content of this relation, which is revealed using a picture. Fig.~\ref{fig:2dbraid}c depicts the expectation value of the relevant product of tunneling operators as a time sequence of events where particle---antiparticle pairs tunnel across the periodic system. Note that the initial and final state in the picture are the same. The four worldline segments in this process, belonging to two quasiparticle types, can be connected to reveal two \textit{linked} worldlines, Fig. ~\ref{fig:2dbraid}d. The picture shows how the $\mathcal{S}$ matrix element and the braiding statistics can be seen in the linking of worldlines. One can detect the linking of worldlines in a purely algebraic way, without drawing the two figures, by calculating the linking topological invariant of the two worldlines. This invariant simply counts the number of links in the one-dimensional worldlines living in the three-dimensional spacetime.

There is another way to relate an $\mathcal{S}$ matrix element to a topological property of the quasiparticle braiding process in 2+1D, which will be important for our generalization to 3+1D. Recently, the connection between modular transformations on the ground state manifold in 2+1D and the statistics of quasiparticles was further exposed by the introduction of minimum entropy states (MES), a special choice of basis in ground state manifold.\cite{Zhang:2012p7534} Namely, an $\mathcal{S}$ matrix element is related to an overlap of two MESs.\cite{Zhang:2012p7534} This relation is due to the fact that every MES can be created by the action of a quasiparticle tunneling operator
% , which is topological since tunneling is across some direction in the periodic 2d system,
on the appropriate reference state.\cite{Zhang:2012p7534} A pictorial representation again uncovers the topological content in this understanding:
% of $\mathcal{S}$ matrix elements and braiding statistics
Fig.~\ref{fig:2dbraid}a depicts the overlap of two MESs as a time sequence of applying quasiparticle tunneling operators. Note that here the initial and final states in the picture are different reference states, being rotated by $90^\circ$.
% is represented as a space-time event of a particle---antiparticle pair tunneling across the periodic 2d system.
In this case we immediately obtain the two worldlines, since each MES in the overlap contributes one. The picture reveals that the two worldlines in this process are linked when the 2+1D space-time process is embedded in three-dimensional space as in Fig. ~\ref{fig:2dbraid}b. This approach again connects the measurement of braiding statistics ($\mathcal{S}$ matrix element) in 2+1D to the linking of particle worldlines.
%Therefore an $\mathcal{S}$ matrix element describes braiding which can be seen as linking of worldlines.

In this paper, we will generalize both above approaches to 3+1D cohomological gauge theory, and obtain some surprising relations between the matrix elements of the three-torus $S,T$ transformations and the braiding of excitations. Noticing that the line, which represents a quasiparticle tunneling operator across a periodic direction on the two-torus, becomes a membrane, which represents the tunneling of a flux-loop across two periodic directions in the three-torus, we will construct the appropriate membrane operators as well as the MES on the three-torus.
%, and truly involves the three dimensional nature of the system.
We will show that in 3+1D topological order, the non-trivial matrix elements of the $S,T$ transformations in the MES basis are due to a non-trivial algebra of the topological membrane operators, analogously to the 2+1D case.
%In this paper, we will calculate the matrix elements of the three-torus $S,T$ transformations in cohomological gauge theory, and relate them to the braiding of excitations.

%This unexpected conclusion is based on our
We will next generalize to 3+1D the two 2+1D approaches described above. Namely, we will study the space-time process representing the non-trivial product of membrane operators, i.e., the loop tunneling operators, as well as the space-time process of overlapping two MESs. Both calculations give the same value, which equals a matrix element of the three-torus $S$ transformation. This value also represents a topological quantum phase accrued during \textit{some} loop space-time process, and the main question becomes: What is the nature of this loop process? In the analogous situation in 2+1D, we described that the particle process is simple braiding, which is characterized by the linking of worldlines.

Most strikingly, we will argue that the obtained $S$ matrix elements in 3+1D relate to certain braiding processes involving \textit{three} loops simultaneously. This is surprising since there is a simple, seemingly fundamental, braiding process of just two loops, where one loop traces out a torus enclosing the other loop, which is relevant in other physical contexts.\cite{Baez:2007p7977,Niemi:2005p7786}

To uncover the topological underpinning of the two studied 3+1D space-time processes, we will consider the topological invariants of the loop worldsheets in both of them. Although we cannot draw the pictures of two-dimensional worldsheets living in the four-dimensional space-time, in analogy with Fig.~\ref{fig:2dbraid}, we will find through calculation that the loop worldsheets in both space-time processes are described by the exact same non-trivial values of the \textit{triple linking number}.\cite{Carter:2001p7981} The triple linking number (TLN) is an integer invariant of closed surfaces in four dimensions, and can be seen as the 3+1D generalization of the linking number of closed lines in three dimensions which was relevant for particles in the 2+1D case.

\begin{figure}
  \includegraphics[width=0.48\textwidth]{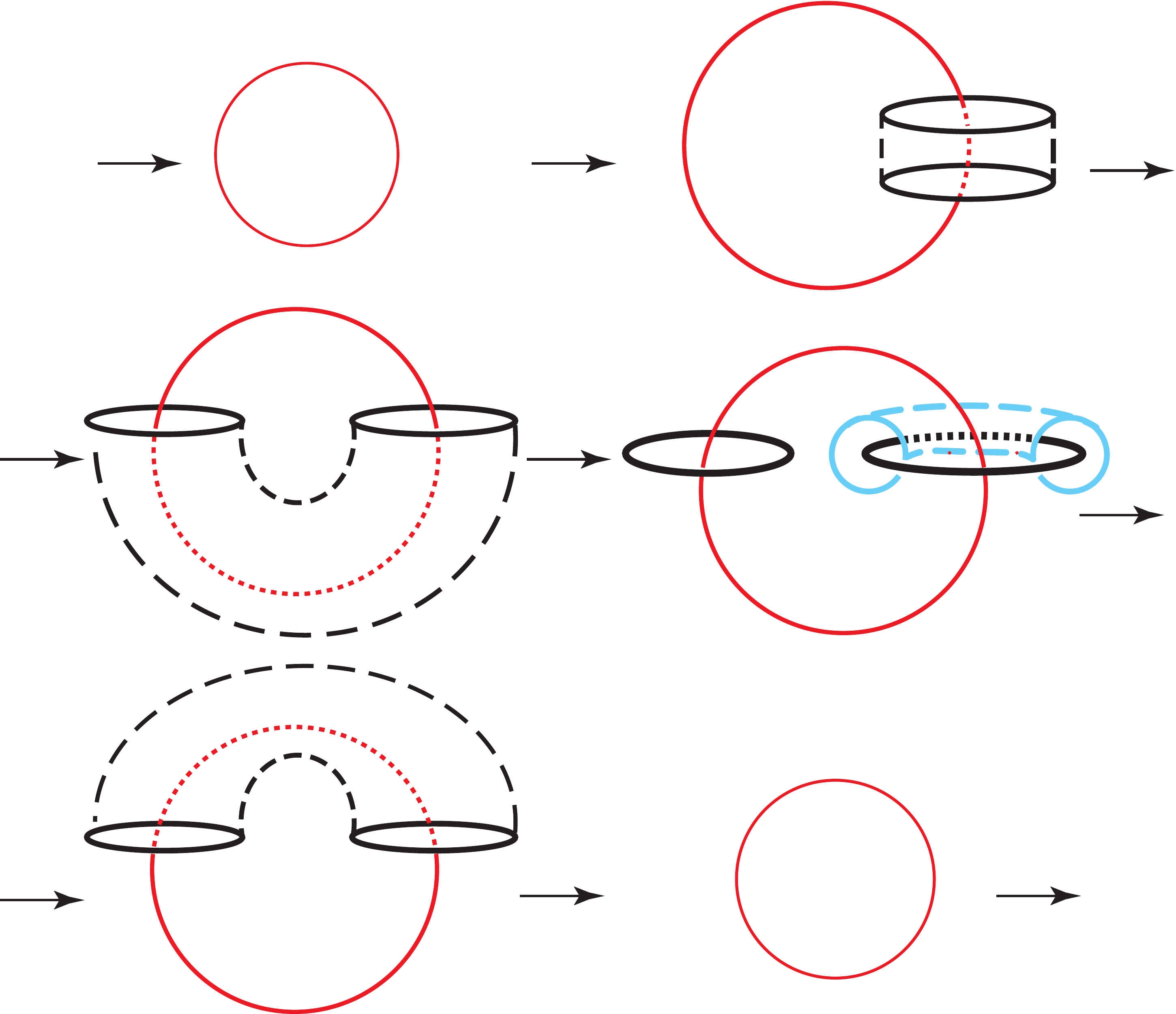}
  \caption{(color online) Movie for three-flux-loop braiding. This process has nontrivial triple linking number of three worldsheets. Firstly, loop-$G$ (red) is created and grows, forming the $G$ worldsheet. Then loop-$H$ (black) emerges, encircling loop-$G$ halfway. Then loop-$F$ (blue) completely encircles loop-$H$. After this, loop-$H$ finishes the route around loop-$G$. Finally, loop-$G$ is annihilated.}
  \label{fig:membrane_linking_movie}
\end{figure}

The TLN is obviously the fundamental, truly three-dimensional descriptor of the relevant three-loop braiding process which is revealed in the context of this paper. To capture its meaning in a tangible way, in this paper we will also construct a movie of a process involving three loops, shown in Fig.~\ref{fig:membrane_linking_movie}, such that the three loop worldsheets in space-time exhibit exactly the same values of TLN as obtained in the two processes above. This movie shows the physical braiding of three flux-loops resulting in a topological quantum phase which measures the properties of the underlying 3+1D topological order (at least for cohomological gauge theory), and equals a matrix element of the three-dimensional $S$ modular transformation.
%We then show that this specific three-loop braiding process is characterized by a non-trivial topological invariant, the \textit{triple linking number},\cite{Carter:2001p7981} of the worldsheets of three loops in the 3+1D spacetime.
%This therefore is the appropriate generalization of situation in 2+1D spacetime, where braiding of particles occurs when particle worldlines, forming closed loops, are non-trivially linked.\cite{Wen:1990p5870}

%Based on models on torus, we extract topological observables of ground state manifold. Further, we identify these observables as braiding statistics and topological spin of topological excitations (including point charge and flux loop). Our work can be viewed as 3+1D extension of previous works [DW, YS Wu, Mesaros Ran].
This paper is organized as follows. In Section~\ref{sec:DW}, we define the exactly solvable models in 3+1D, which are classified by cohomology group and can be viewed as extension of Dijkgraaf-Witten theory to 3+1D. In the following section, we put these models on three-torus, and find the ground state manifolds. Particularly, we find a MES basis, which is useful for interpretation. Further, we construct membrane operators defined as operators mapping between MES. We work out the modular transformations on MES basis in Section~\ref{sec:topo}. We find modular transformations to be directly related to braiding statistics of flux-loops and particles. We show this by both geometric and algebraic methods. In the last Section we solve these models for some illuminating examples.

\section{Cohomological gauge theory in 3+1D}
\label{sec:DW}
In this section, we define the cohomological gauge theory for a general manifold in 3+1D, based on the Dijkgraaf-Witten (DW) topological invariant. The theory is topological and defined by a discrete gauge group $G$. However, there are distinct topologically ordered states for a fixed $G$, and in 3+1D they are classified by the fourth cohomology group of $G$ with coefficients in $U(1)$, namely $H^4(G,U(1))$. In Appendix~\ref{app:coh} we give a brief review of cohomology concepts relevant for the rest of the paper, while referring the reader to, e.g., Refs.\onlinecite{Chen:2013p6670,Mesaros:2013p7698}, for more details.

In this paper we will work in 3+1D, and therefore the theory will be defined using the 4-cocycle (sometimes we call it simply cocycle) $\omega$, for which the cocycle condition becomes:
\begin{align}
  \label{eq:4-cocycle}
  &\omega(g_2,g_3,g_4,g_5)\cdot\omega(g_1,g_2\cdot g_3,g_4,g_5)\cdot\omega(g_1,g_2,g_3,g_4\cdot g_5)\\\notag
  =&\omega(g_1\cdot g_2,g_3,g_4,g_5)\cdot\omega(g_1,g_2,g_3\cdot g_4,g_5)\cdot\omega(g_1,g_2,g_3,g_4),
\end{align}
where $\omega\in H^4(G,U(1))$, and $g_i\in G$. In this paper, we will use the ``canonical'' 4-cocycle, meaning that $\omega(g_1,g_2,g_3,g_4)=1$ if any of $g_1,g_2,g_3,g_4$ is equal to $\mathbf{1}$ (the identity element of group $G$).

\begin{figure}
  \includegraphics[width=0.25\textwidth]{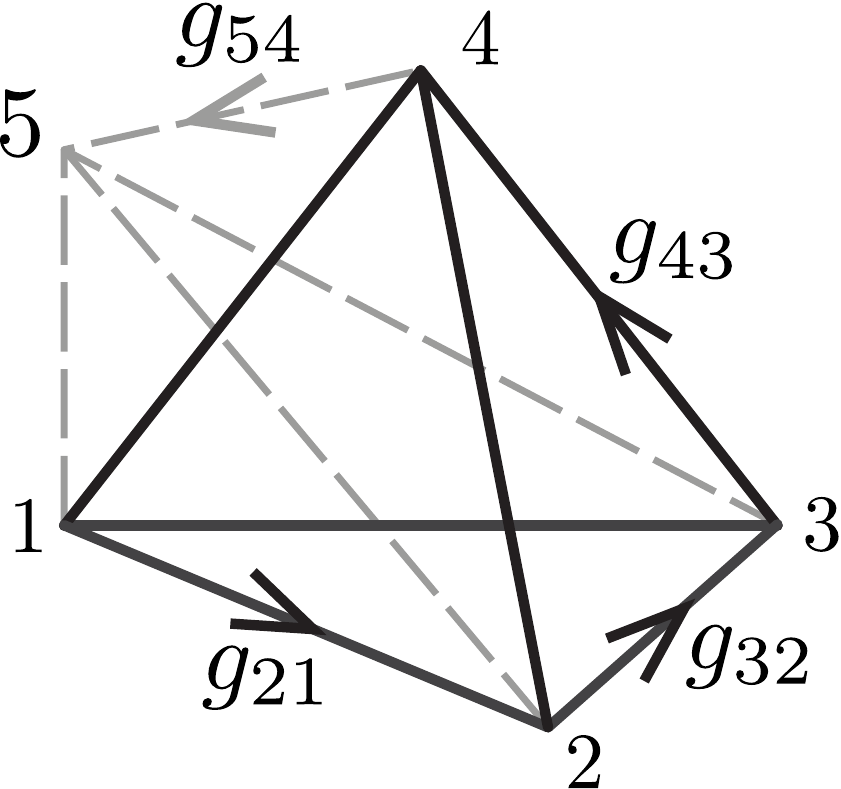}
  \caption{The 4-cocycle $\omega$ assigns a $U(1)$ complex number $\omega^{\varepsilon}(g_{54},g_{43},g_{32},g_{21})$ to a 4-simplex, where $\varepsilon$ is the chirality of the 4-simplex, defined as $\varepsilon=\mathrm{sgn}[\mathrm{det}(\vec{12},\vec{23},\vec{34},\vec{45})]$. The dashed lines represent that the vertex $5$ has a different coordinate in the fourth dimension (time) with respect to the other vertices.}
  \label{fig:4-cocycle}
\end{figure}

The gauge theory is now defined by using $\omega$ to construct topological invariants of a 4D manifold. For a given 4D manifold $M$ without boundary, one can triangulate it using a finite number of 4-simplices.\cite{Nakahara:2003p3627,CoxeterGeometry} The 4-simplex is a higher dimensional analogue of a regular polyhedron, and can be constructed from a tetrahedron by adding a fifth vertex and moving it into the fourth dimension so that all the edges from it to the four original vertices are the same length as the tetrahedron's edges. The triangulation basically requires completely filling the manifold $M$ with 4-simplices without overlaps. The vertices of this triangulation are then ordered arbitrarily, and the ordering is represented by assigning arrows going from the lower to the higher ordered vertex on each edge, Fig.~\ref{fig:4-cocycle}. Let us denote a 4-simplex of the triangulation, together with the ordering of its vertices, by $\sigma_I$, where $I=1,2,\dots,S$ labels 4-simplices and $S$ is the total number of 4-simplices in $M$. Next, one defines a coloring $\varphi$ of all the edges in the triangulation, by assigning group element to them. Let us denote the group element assigned to the bond connecting vertices $j$ and $i$ as $g_{ij}$, following the ordering from $j$ to $i$: $j\rightarrow i$; we then automatically assign $g_{ji}=g_{ij}^{-1}$. In addition, the three assigned group elements for any given face must satisfy the constraint $g_{ij}\cdot g_{jk}\cdot g_{ki}=\mathbf{1}$, and $i,j,k$ are the three vertices of the face. This constraint is the ``zero-flux rule''.

With these definitions, one can assign a $U(1)$ phase to every 4-simplex by computing $\omega^{\varepsilon}(g_{54},g_{43},g_{32},g_{21})$, where $\varepsilon=\mathrm{sgn}[\mathrm{det}(\vec{12},\vec{23},\vec{34},\vec{45}))]$ determines the chirality of the simplex, as shown in Fig. \ref{fig:4-cocycle}.\footnote{The 4D coordinate system $(x,y,z,w)$ itself has a chirality, analogously to the handedness of a 3d coordinate system, and if it changes, the $\varepsilon$ also changes sign.} For a given coloring $\varphi$ and simplex $\sigma_I$, we label this $U(1)$ phase as $W(\sigma_I,\varphi)^{\varepsilon(\sigma_I)}$. Finally, one can compute the product of all $W$ for the simplices: $\prod_{I=1}^S W(\sigma_I,\varphi)^{\epsilon(\sigma_I)}$. For a given coloring $\varphi$, we will have one such product. The key result\cite{Dijkgraaf:1990p7194} is that the complex number
\begin{equation}
 Z_M=\frac{1}{|G|^V}\sum_{\substack{\varphi\in\text{ all}\\\text{possible}\\\text{colorings}}}\prod_{I=1}^S W(\sigma_I,\varphi)^{\epsilon(\sigma_I)},\label{eq:DW_noboundary}
\end{equation}
where $|G|$ is the number of elements in group $G$, and $V$ is the number of vertices in the triangulation, is a topological invariant of the manifold $M$. More precisely, $Z_M$ does not depend on the triangulation and the ordering of vertices (while different colorings are already summed over), owing to the cocycle condition in Eq.~\eqref{eq:4-cocycle}. One can further show that equivalent cocycles (i.e., cocycles differing by a coboundary) give the same value of $Z_M$.\cite{Dijkgraaf:1990p7194}

The topological invariant $Z_M$ is exactly the partition function of the cohomological gauge theory, which is a topological quantum field theory for discrete gauge group $G$ in 3+1D. It is the higher dimensional version of the DW theory\cite{Dijkgraaf:1990p7194,WAKUI:1992p7120}, and it only depends on inequivalent elements in $H^4(G,U(1))$.

\subsection{Exactly solvable models}

We define our exactly solvable models in 3+1D as Hamiltonian versions of the cohomological gauge theory. We consider space triangulated using a tetrahedron lattice with oriented edges (bonds), where these orientations are compatible with some ordering of lattice sites, and assign an element $g_{ij}\in G$ to each oriented edge $j \to i$, according to the above discussion.

An arbitrary quantum state in the Hilbert space $\mathcal{H}$ of our model is then labeled by $|a\rangle=|\{g_{ij}\}\rangle$. The building block for the Hamiltonian is the operator $\hat{B}^\gG_p$ labeled by a group element $\gG\in G$, and a ``plaquette'' $p$ containing all 3-simplices (tetrahedra) that share the vertex $i$. The plaquette operator acts on group elements on the edges that share $i$. To define its action, we introduce an additional edge rising into the fourth dimension, connecting $i$ to an auxiliary vertex $i'$. To edge $i\to i'$ we assign the element $\gG\in G$. The group elements are changed as
\begin{align}
  \label{eq:19}
  g_{ij}&\rightarrow \gG\cdot g_{ij}\\\notag
  g_{ki}&\rightarrow g_{ki}\cdot \gG^{-1},
\end{align}
and these new values are represented on auxiliary edges $i'\to j$ and $k\to i'$. Further, the non-zero matrix elements of $\hat{B}^\gG_p$, namely $B^\gG_p=\langle\text{f}(\gG)|\hat{B}^\gG_p|\text{i}\rangle$, are assigned the following quantum amplitude
\begin{equation}
  \label{eq:2}
  B^\gG_p\equiv\prod_{I=1}^6 W(\sigma_I,\varphi)^{\varepsilon(\sigma_I)},
\end{equation}
where the 4-simplices $\sigma_I$ are built by triangulating the 4D volume formed by the tetrahedra in the plaquette $p$ and the auxiliary edges.
%For example, in Fig.~\ref{fig:4-cocycle} vertex $5$ could represent $i'$.

It is important to note that the zero-flux rule is by construction satisfied on all faces (triangles) of 4-simplices, if it is satisfied in the tetrahedra of $p$, and this must be imposed for the $B^\gG_p$ to be well-defined.
%Finally, note that choosing a final state $|\text{f}\rangle$ in a non-zero matrix element fixes a unique value of $\gG$.
We can then define the plaquette operators $\hat{B}_p$ as having matrix elements
\begin{equation}
  \label{eq:3}
  B_p=\frac{1}{|G|}\sum_{\gG\in G}B^\gG_p.
\end{equation}

The $\hat{B}_p$ are projectors, which can be easily checked using the cocycle property to show $\bra{\text{f}}\hat{B}^\gG_p \hat{B}^{\gG'}_p\ket{\text{i}}=B^{\gG\cdot\gG'}_p$, which then implies $\bra{\text{f}}\hat{B}_p \hat{B}_p\ket{\text{i}}=B_p$. Similarly, it can be shown that the plaquette operators commute, $[B_p,B_{p'}]=0,\quad\forall p,p'$.

Let us also introduce the operator $Q_t$, which projects flux in a triangle $t$ to zero, i.e., it enforces the zero-flux rule.
%In other words, $Q_t$ is non-zero (and equal to 1) only when acting on a triangle $t$ made out of lattice sites $i,j,k$ such that
%\begin{align}
 % \hG_{ij}\cdot\hG_{jk}\cdot\hG_{ki}&=\openone_{\GG}
%\end{align}
%where $\openone$ is the group identity in $\G$, and the second line follows directly from the definition Eq.~(\ref{eq:53}).
Then the Hamiltonian takes the form
\begin{equation}
  \label{eq:5}
  H=-\sum_t Q_t - \sum_p \hat{B}_p\prod_{t\in p}Q_t,
\end{equation}
where the label $t\in p$ enumerates all the triangles making up the plaquette $p$. As mentioned above, the factor $\prod_{t\in p}Q_t$ is actually crucial to ensure that $H$ is well-defined. Further, it is easy to see that plaquette operator term $\hat{B}_p\prod_{t\in p}Q_t$ actually commutes with the projectors $Q_{t'}$.
%Namely, the transformation rule by $\gG$ in operator $\hat{B}^\gG_p$, as introduced above, preserves the product rule Eq.~\eqref{eq:9} on all triangle faces of simplices in Fig.~\ref{fig:1}, if it is satisfied in either the upper or lower triangles, i.e. either in the $\ket{\text{i}}$ or $\ket{\text{f}}$ state. Obviously then the zero-flux rule enforced by action of $Q_{t'}$ commutes with the action of $\hat{B}_p\prod_{t\in p}Q_t$ even when $t'$ belongs to the plaquette $p$.
Since all the terms in $H$ commute with each other, the model is exactly solvable.

Let us briefly mention the connection of the Hamiltonian formulation to the gauge theory, which is exhibited in the ground state manifold. Since all the terms in $H$ are projectors, the ground state manifold is the image of the projector $P=\prod_p\hat{B}_p\prod_{t\in p}Q_t$. On the other hand, $P$ is exactly the projector defining the cohomological gauge theory on the 4D manifold having two copies of our spatial manifold $M$ as boundaries (see Ref.\onlinecite{Mesaros:2013p7698} for details).
% Namely, as $P$ applies all $\hat{B}_p$ operators on $M$, it creates a lifted copy of $M$, leaving the 4D volume $\tilde{M}$ between them triangulated by 4-simplices; the transition amplitude for this operation is equal to the product of $\prod_{I=1}^6 W(\sigma_I,\varphi)^{\varepsilon(\sigma_I)}$ phases, and therefore exactly equals the Dijkgraaf-Witten topological invariant $\tilde{Z}_\tilde{M}$ defined for manifolds with boundaries. (Note that there are no vertices inside the volume, and the number of plaquettes $p$ is equal to $V_{\partial \tilde{M}}/2$ since there are two planes in $\partial M$, leading to correct prefactor from Eq.~(\ref{eq:DW_boundary}).) We can therefore conclude that t
The ground state sector of $H$, to which $P$ projects with eigenvalue 1, is also the ground state sector of the cohomological gauge theory\cite{Dijkgraaf:1990p7194} defined on $M$.
%Actually, it is also easy to see that $\hat{B}_p^s\hat{B}_p=\hat{B}_p$ due to Eq.~\eqref{eq:7} and the group property. This means that for a ground state it also holds that $\hat{B}_p^s=1, \forall p,s$.

\subsection{Geometrical reduction of 4-cocycles}

\begin{figure}
  \includegraphics[width=0.25\textwidth]{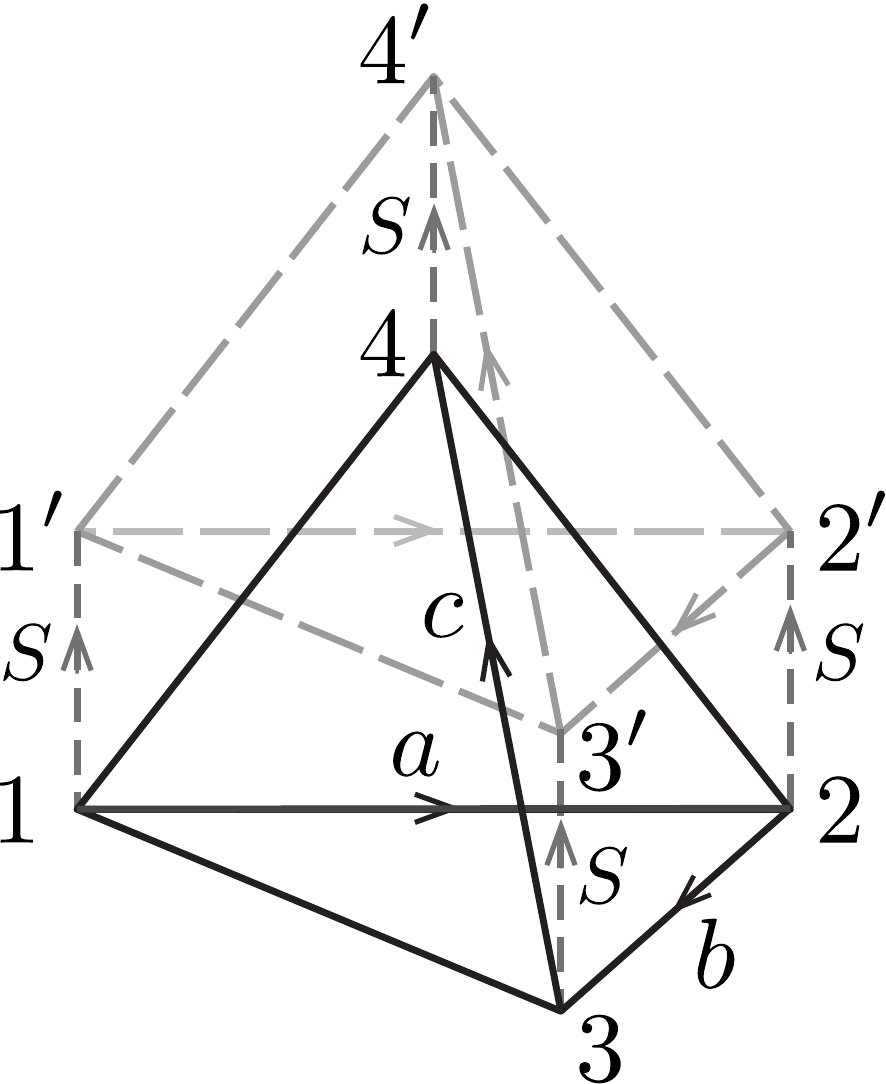}
  \caption{Geometric meaning of 3-cocycle $\beta_s(c,b,a)$ corresponds to evolution (along fourth dimension) of tetrahedron [1234] to [$1'2'3'4'$].}
  \label{fig:phase_for_tetrahedron}
\end{figure}

\begin{figure}
  \includegraphics[width=0.48\textwidth]{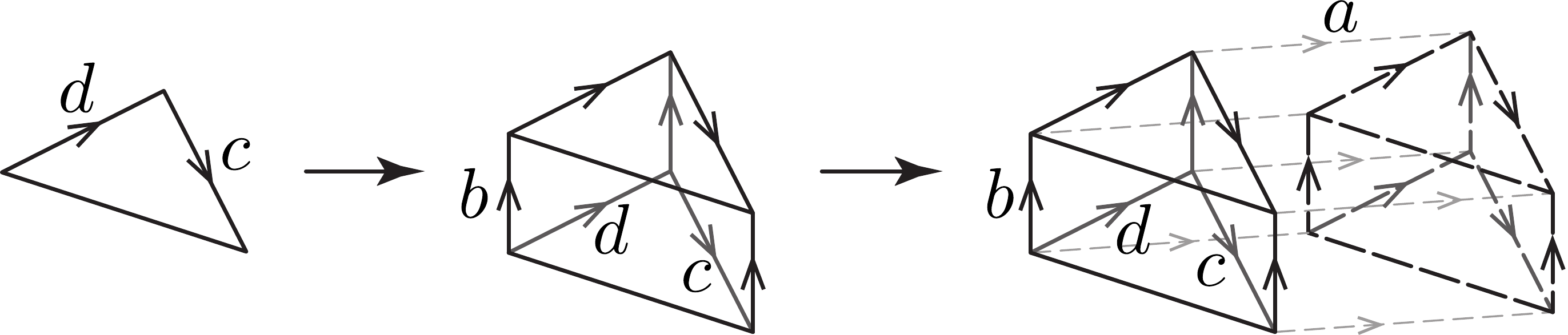}
  \caption{Evolution from a triangle to a 4D manifold. Phase associated with this colored manifold is $\gamma_{a,b}^{\varepsilon}(c,d)$, where $\varepsilon=\mathrm{sgn}[\mathrm{det}(\vec{d},\vec{c},\vec{b},\vec{a}))]$. This phase can also be written as $\gamma_{b,a}^{\varepsilon'}(c,d)$, where $\varepsilon'=\mathrm{sgn}[\mathrm{det}(\vec{d},\vec{c},\vec{a},\vec{b})]=-\varepsilon$. So, we conclude that $\gamma_{a,b}(c,d)=\gamma_{b,a}^{-1}(c,d)$.}
  \label{fig:phase_for_half_cubic}
\end{figure}

In this section we present some cohomology equations for reducing the 4-cocycle to lower order cocycles, and explain their geometric meaning. These equations crucially simplify all following calculations. From now on, we will focus on Abelian groups $G$ for convenience.

First, let us consider a triangulated 4D manifold in Fig.\ref{fig:phase_for_tetrahedron}, with the shown coloring. (Note that some edges needed for full 4D triangulation are omitted, but coloring and ordering are fully defined.) The $U(1)$ phase calculated from all the 4-simplices spanning this 4D volume, with the 4-cocycle $\omega$ given, equals $\beta^\varepsilon_s(c,b,a)$, with:
\begin{align}
  \beta_s(c,b,a)=\frac{\omega(s,c,b,a)\cdot\omega(c,b,s,a)}{\omega(c,s,b,a)\cdot\omega(c,b,a,s)},
  \label{eq:beta_def}
\end{align}
and $\varepsilon=\mathrm{sgn}[\mathrm{det}(\vec{a},\vec{b},\vec{c},\vec{s})]$. Using the 4-cocycle condition for $\omega$, it is straightforward to show that $\beta_s$ is a 3-cocycle. This shows that lifting all vertices of a tetrahedron produces a quantum phase which is only a 3-cocycle, for any given $\omega$.

Another quantity that appears naturally from a cubic geometry is $\gamma_{a,b}$, whose geometric meaning is shown in Fig. \ref{fig:phase_for_half_cubic}. It is defined from the 3-cocycle $\beta_a$ as:
\begin{align}
  \gamma_{a,b}(c,d)=\frac{\beta_a(b,c,d)\beta_a(c,d,b)}{\beta_a(c,b,d)}.
  \label{eq:gamma_def}
\end{align}
It is straightforward to show that $\delta\gamma_{a,b}(c,d,e)=1$, namely, $\gamma_{a,b}$ is a 2-cocycle (see Appendix~\ref{app:coh}). Further, from Eq.(\ref{eq:beta_def}) and Eq.(\ref{eq:gamma_def}), one can show that $\gamma_{a,b}(c,d)=\gamma_{b,a}^{-1}(c,d)$. This equality follows also from the geometry in Fig.~\ref{fig:phase_for_half_cubic}.

\section{Ground state on three-torus and membrane operators}

\subsection{Exact models on three-torus}

\begin{figure}
  \includegraphics[width=0.2\textwidth]{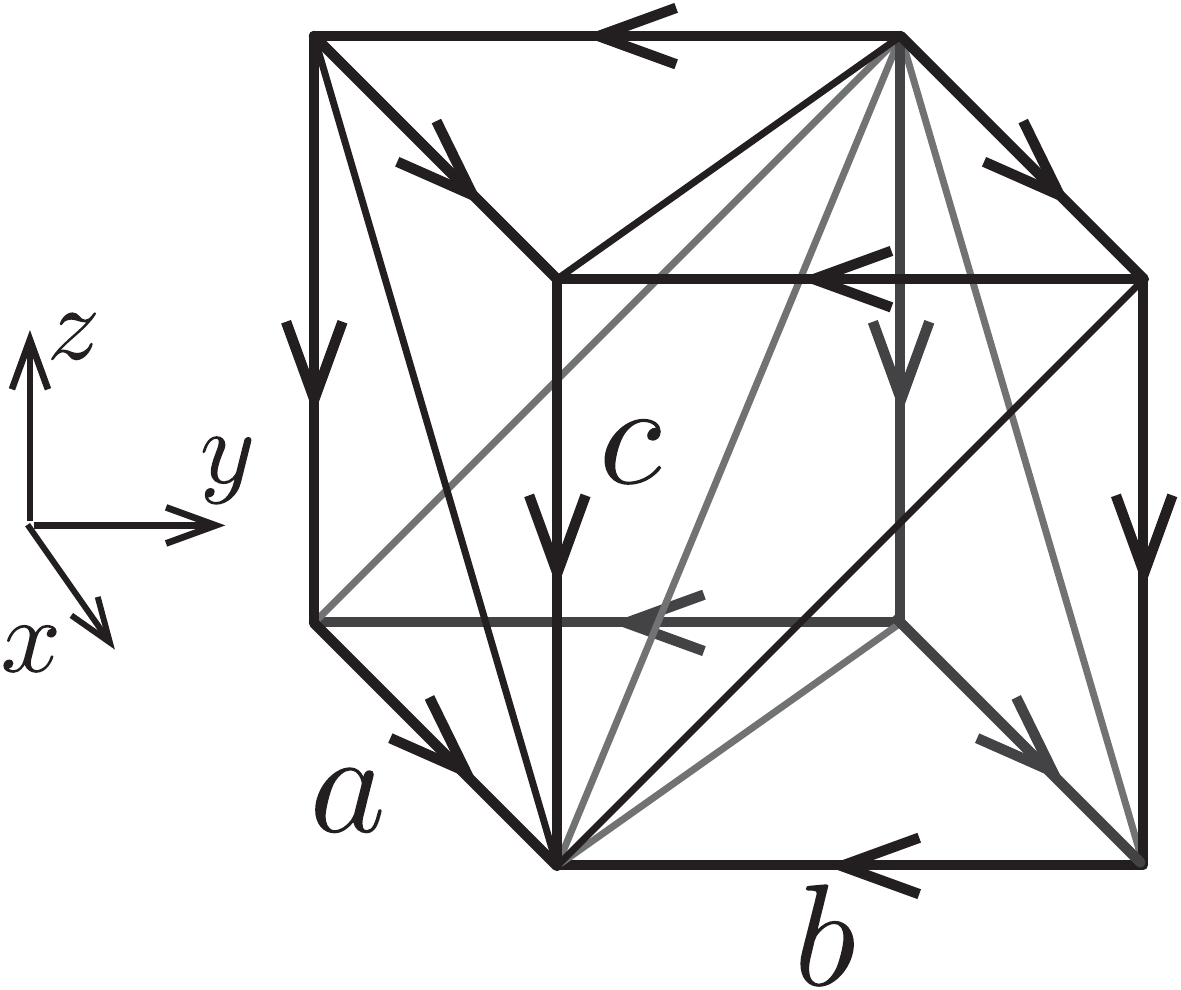}
  \caption{The simplest triangulation of three-torus has a single vertex and three independent edges. Periodic boundary conditions are imposed on the cube.}
  \label{fig:3-torus_triangulation}
\end{figure}

We now put our model, Eq.~\eqref{eq:5}, on the three-torus in 3+1D. It is important to note that the exactly solvable model has correlation length zero. Therefore, we can consider the simplest triangulation of a three-torus shown in Fig.\ref{fig:3-torus_triangulation}. All eight cube vertices are identical due to periodic boundary conditions. It is triangulated by six tetrahedrons.  There are three independent edges, which are assigned group elements $a,b,c\in G$, with $G$ a finite group. Edges with the same direction share the same group element value. The corresponding quantum state is labeled by $|a,b,c\rangle$. We also require $G$ to be Abelian for simplicity.

Since there is only one vertex, we denote the plaqutte operator $\hat{B}_p$ simply as $\hat{B}$, which equals $\frac{1}{|G|}\sum_{s\in G}\hat{B}^s$. The action of $\hat{B}^s$ on state $|a,b,c\rangle$ is
\begin{align}
  \label{eq:Bs_act_3-torus}
  \hat{B}^s|a,b,c\rangle&=\frac{\gamma_{a,s}(b,c)}{\gamma_{a,s}(c,b)}|a,b,c\rangle.\\\notag
  &=\frac{\gamma_{a,b}(c,s)}{\gamma_{a,b}(s,c)}|a,b,c\rangle.
\end{align}
We can directly write down the above result due to the observation that the 4D graph we obtain by acting with $\hat{B}^s$ is in fact made out of two copies of Fig. \ref{fig:phase_for_half_cubic}.\footnote{The precise way in which the $\hat{B}^s$ operator acts in the three-torus with a single vertex (Fig.~\ref{fig:3-torus_triangulation}) is obtained by first adding the (identical) edge rising into the fourth dimension from all eight (identical) corners of the cube; then for each of six tetrahedra forming the cube one obtains the object in Fig.~\ref{fig:phase_for_tetrahedron}. This object can be triangulated into four 4-simplices, giving a total of 24 4-simplices which contribute to the quantum amplitude of $\hat{B}^s$. An insightful shortcut is based on Fig.~\ref{fig:phase_for_half_cubic}, which gives in elegant form the total amplitude contribution coming from half of the cube.} Notice that the $U(1)$ phase obtained by action of $\hat{B}^s$ is a fully antisymmetric function of $a,b,c,s$, as can be seen both geometrically and algebraically.

\subsection{MES as ground state basis}

Let us first briefly review topological order in 2+1D. It is partially characterized by ground state degeneracy on torus.\cite{Wen:1990p5870} One can understand this degeneracy by applying Wilson loop operators of distinct topological excitations winding around one of noncontractible loops on the torus. From this point of view, one can see that ground state degeneracy equals the number of distinct topological superselection sectors.

Nonchiral topological order is fully determined by braiding statistics and topological spin of its topological excitations.\cite{Wen:1992p6753} Remarkably, one can read the information about excitations from ground state by using modular transformations\cite{Wen:1990p7458,Zhang:2012p7534}, namely, by considering the $\mathcal{S,T}$ matrices of the $S,T$ transformation in the ground state manifold.  Dimension of $\mathcal{S,T}$ equals the number of topological sectors. In a proper ground state basis, we can obtain the ``canonical form'' of $\mathcal{S,T}$ matrices, for which the entries of $\mathcal{S}$ matrix are the braiding statistics and the diagonal elements of $\mathcal{T}$ are the topological spins of quasiparticles. Ground state basis for canonical $\mathcal{S,T}$ matrices is formed by minimal entropy state (MES).\cite{Zhang:2012p7534}

We can extend these concepts to 3+1D. However, there is a major difference in this case: Topological excitations can be flux loops in 3+1D. Without loss of generality, we only consider the MES in $z$ direction.

Inspired by the case of 2+1D cohomological gauge theories discussed in Ref.\onlinecite{Hu:2012p7528}, we have found the MES in $z$ direction as
\begin{align}
  |a,b,\lambda\rangle=\frac{1}{\sqrt{|G|}}\sum_{c\in G}\tilde{\chi}_{\lambda}^{a,b}(c)|a,b,c\rangle,
  \label{eq:MES_z}
\end{align}
where $\tilde{\chi}_{\lambda}^{a,b}$ is a one-dimensional projective representation. Here, $\lambda$ labels different projective representations of the group $G$ (see Appendix~\ref{app:coh}), and the 2-cocycle $\gamma$ from Eq.~\eqref{eq:gamma_def} plays the role of factor-system of these projective representations:
\begin{align} \tilde{\chi}_{\lambda}^{a,b}(c_1)\tilde{\chi}_{\lambda}^{a,b}(c_2)=\gamma_{a,b}(c_1,c_2)\tilde{\chi}_{\lambda}^{a,b}(c_1c_2).
  \label{gamma_proj_rep}
\end{align}
\textit{We will only consider the case of Abelian (one-dimensional) projective representations $\chi^{a,b}$ in this paper. This assumption implies that the 2-cocycle $\gamma_{a,b}$ is a 2-coboundary.} We believe this is related to the physical assumption of Abelian statistics of loops. Note that our assumption about $\gamma_{a,b}$ implies that $\gamma_{a,b}(c,s)=\gamma_{a,b}(s,c)$, so that the phase factor in Eq.~\eqref{eq:Bs_act_3-torus} is just identity, and states $|a,b,c\rangle$ for any $a,b,c\in G$ are in the ground state manifold.

%({\color{red} we need to sharply define $\lambda$. Some puzzles: differen $\gamma^{a,b}$ have diffenrent number of proj rep? What is fusion rule of $\lambda$? Learn more about proj rep theory!})

Firstly, we verify that this state is indeed in the ground state manifold. Acting with projection operator $\hat{B}$ on the state, we get
\begin{align}
  \hat{B}|a,b,\lambda\rangle&=\frac{1}{\sqrt{|G|^3}}\sum_{c\in G}\tilde{\chi}_{\lambda}^{a,b}(c)\sum_{s\in G}B^s|a,b,c\rangle \\\notag
  &=\frac{1}{\sqrt{|G|^3}}\sum_c\tilde{\chi}_{\lambda}^{a,b}(c)\cdot\frac{\gamma_{a,b}(c,s)}{\gamma_{a,b}(s,c)}|a,b,c\rangle\\\notag
  &=\frac{1}{\sqrt{|G|}}\sum_c\tilde{\chi}_{\lambda}^{a,b}(c)|a,b,c\rangle\\\notag
  &=|a,b,\lambda\rangle,
\end{align}
where the second row uses Eq.(\ref{eq:Bs_act_3-torus}), and in the third row we used $\gamma_{a,b}(c,s)=\gamma_{a,b}(s,c)$ which follows from the above mentioned assumptions.
%({\color{red} This assumption is not generally true even when $G$ is abelian. When proj rep is non-Abelian, I don't have proof that the GS defined above is right. From now on, we only consider Abelian group G and Abelian proj rep}).

\begin{figure}
  \includegraphics[width=0.2\textwidth]{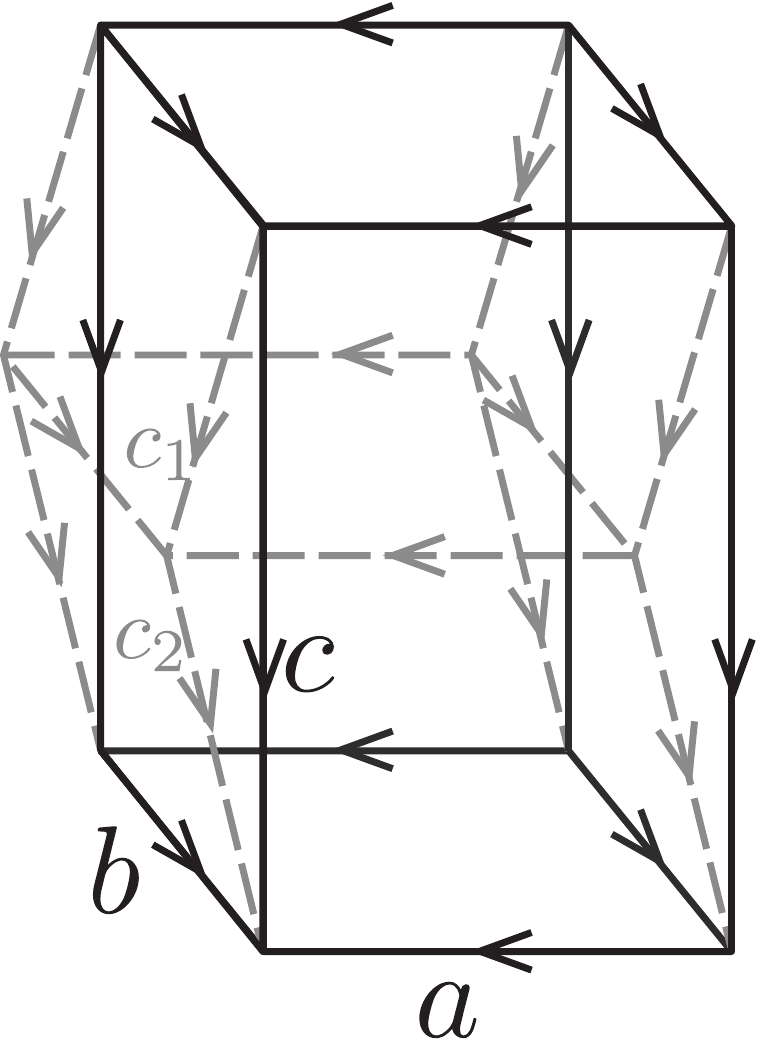}
  \caption{Evolution from single vertex to two vertices.}
  \label{fig:3-torus_2uc}
\end{figure}

Next, we prove that this state is indeed an MES in $z$ direction. Let us retriangulate the three-torus, so that it has two unit-cells in $z$ direction. The ground state defined on this two unit-cell system can be evolved from that in one unit-cell, as shown in Fig.~\ref{fig:3-torus_2uc}:
\begin{align}
  \label{eq:MES_z_2uc}
  |a,b,\lambda\rangle&=\frac{1}{\sqrt{|G|}}\sum_{c\in G}\tilde{\chi}_{\lambda}^{a,b}(c)|a,b,c\rangle\\\notag
  &=\frac{1}{\sqrt{|G|}}\sum_{c_1,c_2\in G}\tilde{\chi}_{\lambda}^{a,b}(c_2\cdot c_1)\gamma_{a,b}(c_2,c_1)|a,b,c_1,c_2\rangle\\\notag
  &=\frac{1}{\sqrt{|G|}}\sum_{c_1}\tilde{\chi}_{\lambda}^{a,b}(c_1)\sum_{c_2}\tilde{\chi}_{\lambda}^{a,b}(c_2)|a,b,c_1,c_2\rangle.
\end{align}
As seen from the above, $|a,b,\lambda\rangle$ defined on two unit-cells can be written as a direct product state. So, entanglement entropy of this state in $z$ direction is zero, which must be minimum. We therefore conclude that this state is indeed an MES in $z$ direction. Note that the entanglement entropy for this small system size vanishes due to a cancellation between the ``area law'' contribution and the topological contribution (see Ref.\onlinecite{Zhang:2012p7534} and references therein), which will not occur for larger system sizes.

Similarly, it is easy to write down the MES in $x$ and $y$ directions:
\begin{align}
  \label{eq:MES_x}
  |\mu,b,c\rangle=\frac{1}{\sqrt{|G|}}\sum_{a\in G}\tilde{\chi}^{b,c}_{\mu}(a)|a,b,c\rangle,\\
  \label{eq:NES_y}
  |a,\nu,c\rangle=\frac{1}{\sqrt{|G|}}\sum_{b\in G}\tilde{\chi}^{c,a}_{\nu}(b)|a,b,c\rangle,
\end{align}
whose properties can be derived in the same way as above.

%({\color{red} prove number of MES equals to GSD?})

\subsection{Membrane operator}

\begin{figure}
  \includegraphics[width=0.48\textwidth]{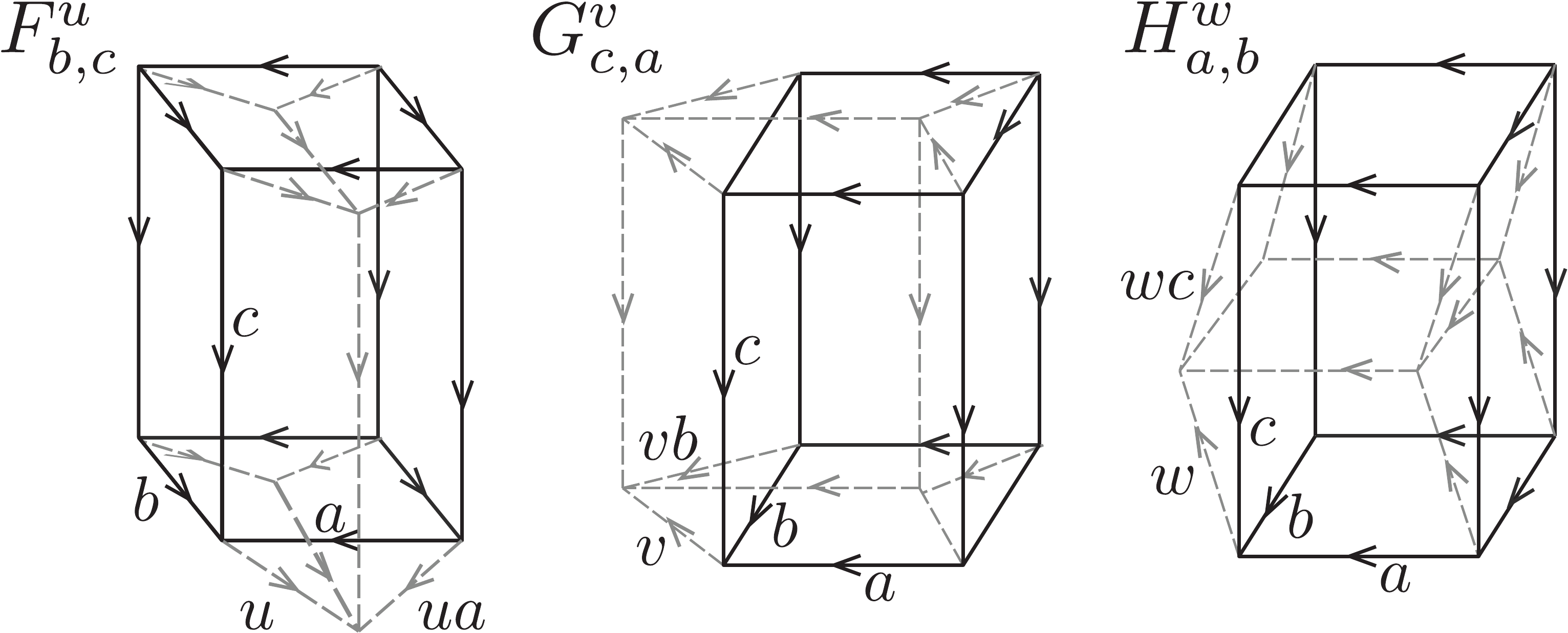}
  \caption{The action of membrane operators.}
  \label{fig:membrane_operator}
\end{figure}

Although we constructed the MES in 3+1D, the physical picture is still unclear. Recall that in 2+1D all MES can be obtained from inserting ribbon operators (Wilson loop operators) into ``trivial'' MES, which corresponds to topological trivial sector. In the following we will show that membrane operators are the relevant operators for such a procedure in 3+1D.

Let us start with the MES in $z$ direction, $|a,b,\lambda\rangle$. Characteristically in discrete gauge theory, we can interpret a group element as a label of flux-loop. Consequently, a membrane, which is the 3+1D analogue of the Dirac string, is also labeled by a group element. Further, a group representation labels a particle.\cite{Propitius:1993CS_ccy}
% Particle $\lambda$ going around flux loop $g$ will pick up the phase $\chi_\lambda(g)$.
Then $|a,b,\lambda\rangle$ can be viewed as state with membrane $a$ in $yz$ plane and membrane $b$ in $zx$ plane, as well as string $\lambda$ (world-line of particle) in $z$ direction. So, it is natural to define trivial MES as
\begin{align}
  \label{eq:MES_z_trivial}
  |e,e,\mathbbm{1}\rangle=\frac{1}{\sqrt{|G|}}\sum_{c\in G}|e,e,c\rangle,
\end{align}
where $e\in G$ is identity element. Here $\mathbbm{1}$ means the trivial linear representation.

The central question becomes: What are the operators that send one MES to another? It is natural to assume that these operators correspond to membrane insertion in $yz$ and $zx$ plane, as well as string insertion in $z$ direction. Besides, we expect that a string in $x$($y$)  direction can measure a membrane in $yz$($xz$) plane while membrane in $xy$ plane will measure strings in $z$ direction.

Following this intuition, we define membrane insertion operators in $yz$, $zx$, $xy$ planes, respectively, as shown in Fig.~\ref{fig:membrane_operator}:
\begin{align}
  \label{eq:membrane_operator}
  F_{b',c'}^u|a,b,c\rangle&=\delta_{bb'}\delta_{cc'}\cdot\gamma_{b,c}^{-1}(u,a)|ua,b,c\rangle,\\\notag
  G_{c',a'}^v|a,b,c\rangle&=\delta_{cc'}\delta_{aa'}\cdot\gamma_{c,a}^{-1}(v,b)|a,vb,c\rangle,\\\notag
  H_{a',b'}^w|a,b,c\rangle&=\delta_{aa'}\delta_{bb'}\cdot\gamma_{a,b}^{-1}(w,c)|a,b,wc\rangle,
\end{align}
where $u,v,w$ label the spatial planes of the membranes. Further, we can define
\begin{align}
  \label{eq:membranesFG}
  F^{(z)}_{u,\lambda}&=\sum_{b,c\in G}\tilde{\chi}_{\lambda}^{u,b}(c)F_{b,c}^u,\\\notag
  G^{(z)}_{v,\lambda}&=\sum_{c,a\in G}\tilde{\chi}_{\lambda}^{a,v}(c)G_{c,a}^v,
\end{align}
where we interpret $F^{(z)}_{u,\lambda}$ as inserting membrane $u$ (in $yz$ plane) and string $\lambda$ in $z$ direction, and interpret $G^{(z)}_{v,\lambda}$ as inserting membrane $v$ (in $zx$ plane) and string $\lambda$ in $z$ direction. To confirm this, we act with these operators on state $|e,e,\mathbbm{1}\rangle$, getting
\begin{align}
  F^{(z)}_{u,\lambda}|e,e,\mathbbm{1}\rangle&=|u,e,\lambda\rangle\\\notag
  G^{(z)}_{v,\lambda}|e,e,\mathbbm{1}\rangle&=|e,v,\lambda\rangle.
\end{align}

It is not hard to obtain the ``fusion rule'' of membranes and strings, namely
\begin{align}
  \label{eq:same_membrane_fusion}
  F_{u_1,\lambda_1}^{(z)}F_{u_2,\lambda_2}^{(z)}&=F_{u_1u_2,\lambda_3}^{(z)},\\\notag
  G_{v_1,\lambda_1}^{(z)}G_{v_2,\lambda_2}^{(z)}&=G_{v_1v_2,\lambda_3}^{(z)},
\end{align}
and
\begin{align}
  \label{eq:diff_membrane_fusion}
  F_{u,\lambda_1}^{(z)}G_{v,\lambda_2}^{(z)}|a,b,\lambda\rangle&=|ua,vb,\lambda_3\rangle\\\notag
  G_{v,\lambda_1}^{(z)}F_{u,\lambda_2}^{(z)}|a,b,\lambda\rangle&=|ua,vb,\lambda_3\rangle,
\end{align}
where the representation $\lambda_3$ is determined as the one in which every element $g\in G$ is represented by the product of numbers which represent $g$ in the $\lambda_1$ and $\lambda_2$ representations (note that the Abelian representations considered here are always one-dimensional, i.e., just numbers). The fusion rules follow from the properties of the 2-cocycle $\gamma$, namely, assume $\tilde{\chi}_{\mu}^{a,b}$ is a projective representation with factor system $\gamma_{a,b}$,
\begin{align} \tilde{\chi}_{\mu}^{a,b}(c_1)\cdot\tilde{\chi}_{\mu}^{a,b}(c_2)=\gamma_{a,b}(c_1,c_2)\cdot\tilde{\chi}_{\mu}^{a,b}(c_1\cdot c_2).
\end{align}
Then it follows that
\begin{align}
  \label{eq:proj_rep_fusion} \tilde{\chi}_{\mu_1}^{a,b_1}(c)\tilde{\chi}_{\mu_2}^{a,b_2}\gamma_{a,c}(b_1,b_2)&=\tilde{\chi}_{\mu_3}^{a,b_1b_2}(c)\\\notag \tilde{\chi}_{\mu_1}^{a_1,b}(c)\tilde{\chi}_{\mu_2}^{a_2,b}\gamma_{c,b}(a_1,a_2)&=\tilde{\chi}_{\mu_3}^{a_1a_2,b}(c),
\end{align}
where the representations $\mu_1,\mu_2,\mu_3$ are related in the same way as $\lambda_1,\lambda_2,\lambda_3$ just above.
%({\color{red} Sharply define $\mu_3$})
%({\color{red} Notice that $\lambda_3$ we wrote above is in fact not the same one. We need to work out this fusion algebra more carefully}.)

Similarly to the above derivations, we can define
\begin{align}
    \label{eq:membranesHH}
  H_{w,\mu}^{(x)}&=\sum_{a,b}\tilde{\chi}_{\mu}^{b,w}(a)H_{a,b}^w,\\\notag
  H_{w,\mu}^{(y)}&=\sum_{a,b}\tilde{\chi}_{\nu}^{w,a}(b)H_{a,b}^w,
\end{align}
where $H^{(x)}$($H^{(y)}$) creates membrane in $xy$ plane and string in $x$($y$) direction. Acting with these operators on MES in $z$ direction, we get
\begin{align} H_{w,\mu}^{(x)}|a,b,\lambda\rangle&=\frac{\tilde{\chi}_{\mu}^{b,w}(a)}{\tilde{\chi}_{\lambda}^{a,b}(w)}|a,b,\lambda\rangle,\\\notag
  H_{w,\nu}^{(y)}|a,b,\lambda\rangle&=\frac{\tilde{\chi}_{\nu}^{w,a}(b)}{\tilde{\chi}_{\lambda}^{a,b}(w)}|a,b,\lambda\rangle.
\end{align}
It is then natural to interpret $H^{(x)}$($H^{(y)}$) as operator that measures strings in $z$
% ($Z$)
direction and membrane in $yz$($zx$) plane.

We will also write down the remaining two operators that send MES to MES for later convenience:
\begin{align}
    \label{eq:membranesFGmapMES}
  F_{u,\nu}^{(y)}&=\sum_{b,c}\tilde{\chi}_{\nu}^{c,u}(b)F_{b,c}^{u},\\\notag
  G_{v,\mu}^{(x)}&=\sum_{c,a}\tilde{\chi}_{\mu}^{v,c}(a)G_{c,a}^{v}.
\end{align}

\section{Topological observables and their physical interpretation}
\label{sec:topo}

\subsection{$\mathcal{S}$ and $\mathcal{T}$ matrices from modular transformations}

In this section, we will calculate the Berry phase of ground states obtained during modular transformations. The derivation is largely a higher dimensional generalization of 2+1D case in Ref.\onlinecite{Hu:2012p7528}.

In real space, we can write the modular transformations, Fig.~\ref{fig:2dST}b, as
\begin{align}
  \mathcal{S}=
  \begin{pmatrix}
    0 & 0 & 1 \\
    1 & 0 & 0 \\
    0 & 1 & 0 \\
  \end{pmatrix},
  \mathcal{T}^{31}=
  \begin{pmatrix}
    1 & 0 & 0 \\
    0 & 1 & 0 \\
    1 & 0 & 1 \\
  \end{pmatrix}.
\end{align}
The question is what is the action of $\mathcal{S}$ and $\mathcal{T}$ on our exact models? We follow the strategy of Ref.\onlinecite{Hu:2012p7528}, but generalize it to 3+1D. We consider a $T^3\times[0,1]$ manifold ($T^3$ is three-torus), and put the initial ground state at $T^3\times{0}$, final state at $T^3\times{1}$. Then we carefully triangulate the 4D manifold $T^3\times[0,1]$ and compute the quantum amplitude from the initial to the final state. After lengthy but straightforward calculations, we find:
\begin{align}
  \label{eq:ST_action}
  S|a,b,c\rangle&=|b,c,a\rangle\\\notag
  T^{31}|a,b,c\rangle&=\beta_b^{-1}(a,a^{-1}c,a)|a,b,a^{-1}c\rangle.
\end{align}
%({\color{red} draw picture of initial state and final state? Doing triangulationa and calculate phase explicitly?}).

Now we act by $T^{31}$ on MES in $z$ direction:
\begin{align}
  \label{eq:T_act_MES}
  T^{31}|a,b,\lambda\rangle&=\frac{1}{\sqrt{|G|}}\sum_c \tilde{\chi}_{\lambda}^{a,b}(c)\beta_b^{-1}(a,a^{-1},a)|a,b,a^{-1}c\rangle\\\notag
  &=\tilde{\chi}_{\lambda}^{a,b}(a)|a,b,\lambda\rangle.
\end{align}
We can see that $|a,b,\lambda\rangle$ is indeed an eigenstate of $\mathcal{T}$ matrix.

We can also get $\mathcal{S}$ matrix element in $z$-direction MES basis:
\begin{align}
  \label{eq:S_act_MES}
  &\langle a',b',\lambda'|S|a,b,\lambda\rangle\\\notag
  &=\frac{1}{|G|}\sum_{c,c'}\frac{\tilde{\chi}_{\lambda}^{a,b}(c)}{\tilde{\chi}_{\lambda'}^{a',b'}(c')}\langle a',b',c'|b,c,a\rangle\\\notag
  &=\frac{1}{|G|}\frac{\tilde{\chi}^{a,b}_{\lambda}(b')}{\tilde{\chi}_{\lambda'}^{a',b'}(a)}\cdot \delta_{a'b}.
\end{align}
Taking into account our assumption that $\gamma_{a,b}$ is a 2-coboundary, the projective representation $\tilde{\chi}$ can be rewritten as $\tilde{\chi}_{\mu}^{ab}(g)=\varepsilon_{a,b}(g)\cdot\chi_{\mu}(g)$, where $\chi_{\mu}(g)$ is an ordinary linear representation of $G$, and $\varepsilon_{a,b}$ is a 1-cocycle for which $\gamma_{a,b}=\delta\varepsilon_{a,b}$ (see Appendix~\ref{app:coh}). Then we get a factorized form:
\begin{equation}
  \label{eq:S_act_MESfact}
\langle a',b',\lambda'|S|a,b,\lambda\rangle=\frac{1}{|G|}\frac{\chi^{a,b}_{\lambda}(b')}{\chi_{\lambda'}^{a',b'}(a)}\cdot\frac{\varepsilon_{a,b}(b')}{\varepsilon_{a',b'}(a)} \cdot\delta_{a'b}.
\end{equation}

While the physical meaning of this element is not so clear for general case, it is instructive to see the simple case where $a'=b=e$. Then the 1-cocycle part of Eq.(\ref{eq:S_act_MESfact}) is trivial and only $\chi_{\lambda}(b')/\chi_{\lambda'}(a)$ is left. We can interpret this phase as Aharonov-Bohm phase of particles going around a flux loop in three dimensions, namely, particle $\lambda$ sees flux loop $b'$ and particle $\lambda'$ sees flux loop $a$. In the following, we will show how the most general form of $\mathcal{S}$ matrix element, including the 1-cocycle contribution, can be interpreted as statistics of flux loops as well as particles.

\subsection{Braiding statistics from $\mathcal{S}$ matrix and triple linking number}

In 2+1D space-time dimensions, the $\mathcal{S}$ matrix elements can be directly related to quasiparticle braiding by considering quasiparticle tunneling operators.\cite{Wen:1990p5870,Zhang:2012p7534} One may ask is it possible to capture the $\mathcal{S}$ matrix elements in 3+1D space-time dimensions using a loop braiding process by considering membrane operators? We will show that this is indeed possible.

Let us review the 2+1D case, mentioned in the Introduction. An $\mathcal{S}$ matrix element can be expressed as an overlap of two minimum entropy states (MES), where every MES can be created by the action of a quasiparticle tunneling operator on the appropriate reference state.\cite{Zhang:2012p7534} Fig.~\ref{fig:2dbraid}a depicts the MES overlap as a time sequence where application of tunneling operator is represented as a space-time event of particle---antiparticle pair tunneling across the periodic system. We get two worldlines, since both MESs in the overlap contribute one. The occurrence of braiding can be revealed by realizing that the two worldlines in this process are linked when the 2+1D space-time process is embedded in three dimensional space as in Fig. ~\ref{fig:2dbraid}b. Therefore an $\mathcal{S}$ matrix element describes braiding which can be seen as linking of worldlines.

We now present an analogous interpretation of an $\mathcal{S}$ matrix element in 3+1D, using membrane operators and a process involving a triple linking of worldsheets in 3+1D. Starting from the $\mathcal{S}$ matrix element in the MES basis in $z$ direction, Eq.~\eqref{eq:S_act_MES}, it is straightforward to show that:
\begin{align}
  \label{eq:S_matrix_MES_z}
\langle v,w,\mu|S|u,v,\lambda\rangle=&\langle\mu,v,w|u,v,\lambda\rangle=\\\notag
=&\frac{1}{|G|}\frac{\tilde{\chi}_{\lambda}^{u,v}(w)}{\tilde{\chi}_{\mu}^{v,w}(u)}=\\\notag =&\langle1,e,e|(G_{v,1}^{(x)})^{-1}(H_{w,\mu}^{(x)})^{-1}F_{u,\lambda}^{(z)}G_{v,1}^{(z)}|e,e,1\rangle,\\\notag
\end{align}
where the last row follows from the definition of membrane operators. In contrast with the 2+1D case, each MES is created by the action of \textit{two} membrane operators on the reference state $|e,e,1\rangle$. A membrane operator, such as Eq.~\eqref{eq:membranesFG},~\eqref{eq:membranesHH},~\eqref{eq:membranesFGmapMES}, can be interpreted as a space-time event of tunneling a flux-loop---antiflux-loop pair across a plane in the periodic system; for example, from Eq. ~\eqref{eq:membranesFG}, the $F_{u,\lambda}^{(z)}$ describes the tunneling of loop in $zy$ plane. (A membrane also contains a string in its plane, which represents particle tunneling.)

The key question now becomes: What is a robust characterization of the process involving the four membrane operators in Eq.~\eqref{eq:S_matrix_MES_z}? Inspired by the 2+1D case, where embedding the process involving one-dimensional worldlines in three dimensions revealed their linking, we embed the process involving two-dimensional worldsheets into four dimensions. We find that three worldsheets have a non-trivial triple linking number (TLN), which is a topological invariant generalizing the linking number used for worldlines in 2+1D. The details of embedding the 3+1D space-time process into four dimensional space are in Appendix~\ref{app:geom}.

Here we briefly summarize the triple linking number (TLN) invariant.

%\subsection{The triple linking number}

\begin{figure}
  \includegraphics[width=0.48\textwidth]{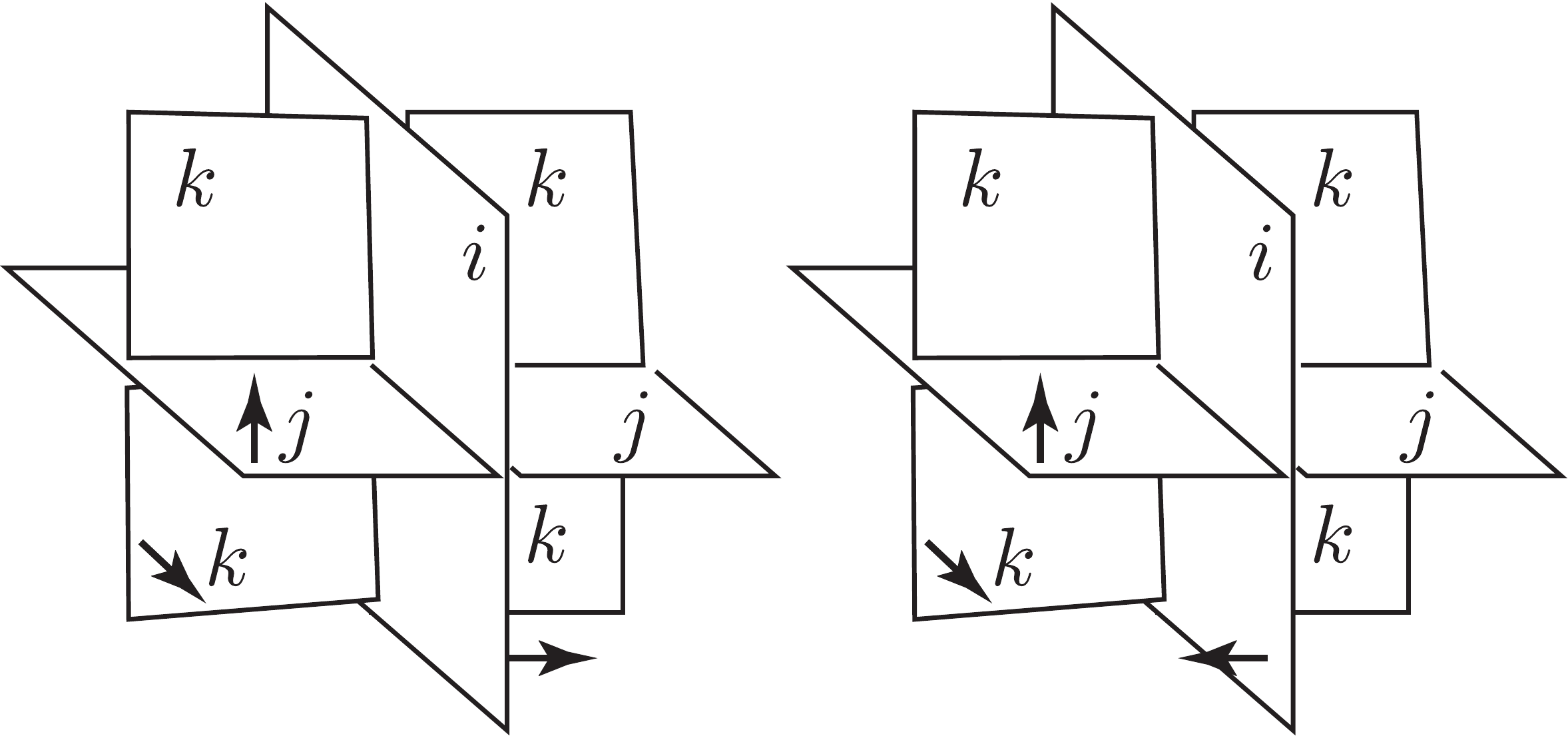}
  \caption{A positive triple point (left) and a negative triple point (right), where we denote the orientations of sheets by their normals.}
  \label{fig:triple_point}
\end{figure}

The triple linking number $Tlk_{MNP}$ of three oriented two-dimensional surfaces $M,N,P$ smoothly embedded in four dimensions was defined in Ref.\onlinecite{Carter:2001p7981} as an analogue of the linking number of classical links.
% The indices $i,j,k$ label three disconnected components of the surface $M$.
In our case, $M,N,P$ are the three flux-loop worldsheets in 3+1D.

$Tlk_{MNP}$ is an integer topological invariant.\cite{Carter:2004p7996} It can be non-zero only if the surfaces $M,N,P$ are distinct, and the $Tlk$ obey the relations
\begin{align}
    \label{eq:tlk}
  Tlk_{MNP}+Tlk_{NPM}+Tlk_{PMN}&=0,\\\notag
  Tlk_{MNP}+Tlk_{PNM}&=0,
\end{align}
and are therefore fully determined by two integers. \cite{Carter:2004p7996}
%Concretely, we can choose:
%\begin{align}
%  \label{eq:1}
%  A&\equiv Tlk_{MNP}\\\notag
%  B&\equiv Tlk_{MPN},
%\end{align}
%which implies $Tlk_{PNM}=-A$, $Tlk_{NPM}=-B$, $Tlk_{NMP}=A-B$, $Tlk_{PMN}=B-A$.

There are different ways to calculate the TLN.\cite{Carter:2004p7996} We describe the one that is most convenient for the braiding problem: One projects the surfaces $M,N,P$ from 3+1D onto a three-dimensional slice using an arbitrary projection direction, and looks for triple-points, namely, points in the projected manifold where all three projected surfaces intersect. For each triple point $s$ one checks the stacking order of surfaces along the projection vector, and assigns the top surface to $I_s$, the middle to $J_s$, and the bottom to $K_s$. Finally, the sign $\epsilon_s$ is calculated as the handedness of the three $I_s,J_s,K_s$ surface normals at the point $s$, see Fig.~\ref{fig:triple_point}. Now let $(I,J,K)$ be a permutation of the three surfaces $(M,N,P)$. Having the above information, $Tlk_{IJK}$ equals the sum of $\epsilon_s$ over the points $s$ for which $I_s=I, J_s=J,K_s=K$. If no triple point contributes to a certain choice $IJK$, then $Tlk_{IJK}=0$, and this has to be consistent with other values of $I'J'K'$ according to relations Eq.~\eqref{eq:tlk}.

The number of triple points and the stacking order of surfaces both depend on the chosen projection vector in 3+1D, however, the resulting TLN is topologically invariant.

%({\color{red} Add figure for triple linking number of S matrix?})

Here we present the TLN result, leaving the calculation details to Appendix \ref{app:geom}. We find that in the embedded process the worldsheets corresponding to membrane operators $F_{u,\lambda}^{(z)}$, $(H_{w,\mu}^{(x)})^{-1}$ and $(G_{v,1}^{(x)})^{-1}$ have triple linking with
\begin{align}
  \label{eq:TLN_S_matrix}
  Tlk_{FHG}&=Tlk_{HFG}=1,\\\notag
  Tlk_{GHF}&=Tlk_{GFH}=-1,\\\notag
  Tlk_{FGH}&=Tlk_{HGF}=0.
\end{align}

{\it We note that even for general 3+1D topological order, beyond cohomological models, the S-matrix elements are given by MES overlaps, while the direct connection between MES overlap and triple linking of tunneling operator worldsheets, such as discussed above and demonstrated using Eq.~\eqref{eq:S_matrix_MES_z}, remains a general property based on the purely geometrical embedding construction.} The only required ingredient is that the MESs of the topological order can be expressed using tunneling operators of loop-like excitations, which we expect to be generally true in 3+1D.

\subsection{Braiding statistics from membrane operator algebra}
\label{sec:braid_memb}
\begin{figure}
  \includegraphics[width=0.48\textwidth]{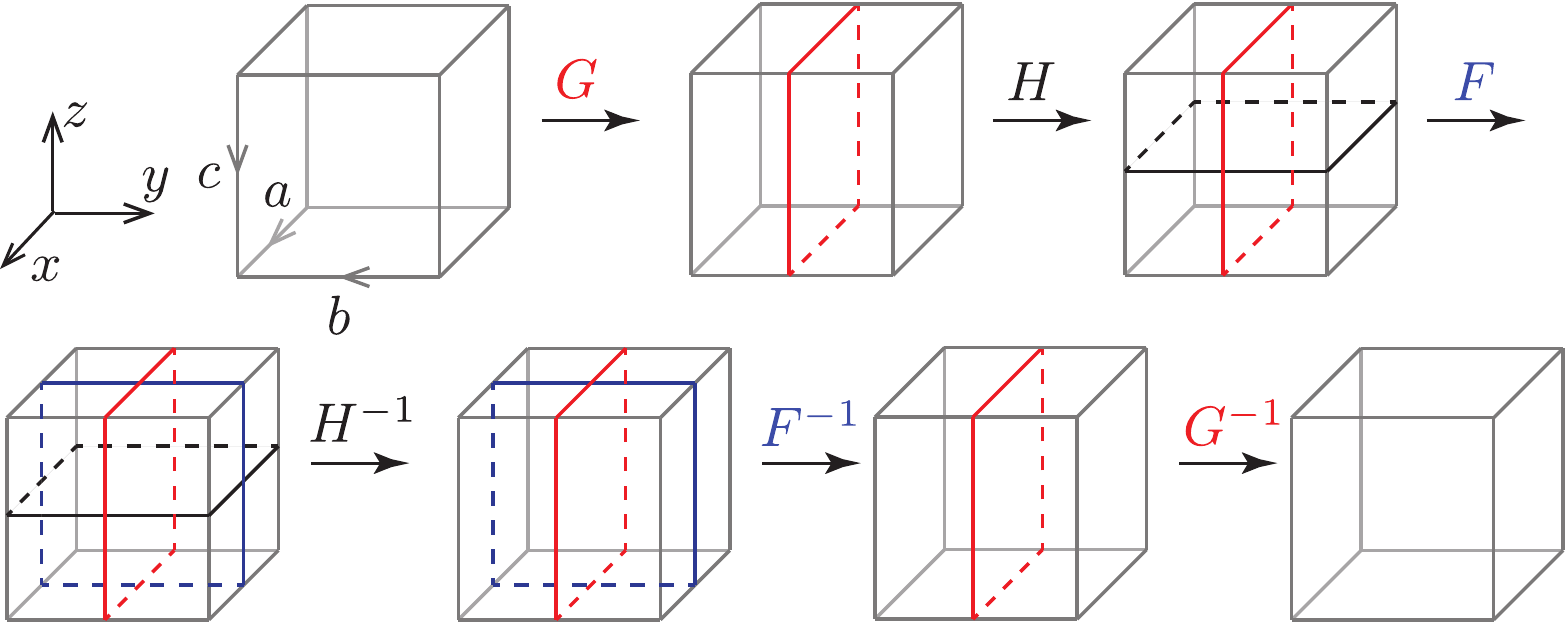}
  \caption{Movie for process $\langle G^{-1}F^{-1}H^{-1}FHG\rangle$. The worldsheets in this process share the same topological properties as the three-flux-loop braiding process in Fig.~\ref{fig:membrane_linking_movie}.}
  \label{fig:membrane_operator_commutation_movie}
\end{figure}

\begin{figure}
  \includegraphics[width=0.25\textwidth]{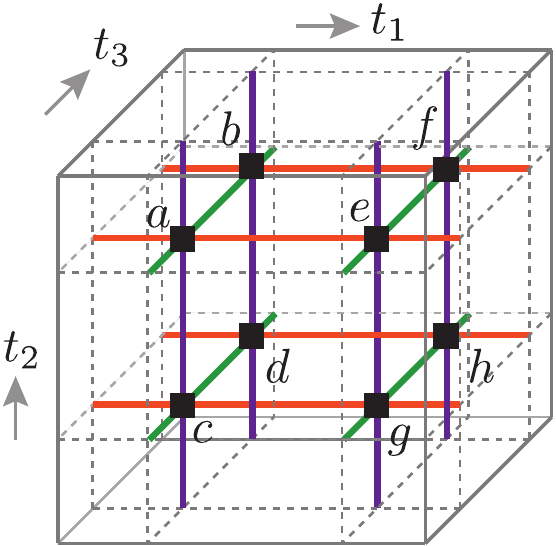}
  \caption{(color online) Projection of the membrane process movie in Fig.~\ref{fig:membrane_operator_commutation_movie} to three-dimensional space at $t=-\infty$. Lines show the pairwise intersections of projected worldsheets. Purple lines: for $F$ and $G$ worldsheets; Orange: for $F$ and $H$; Green: for $G$ and $H$. Although there are eight triple points here, the triple linking is still the same as for the three-flux-loop braiding process, Fig.~\ref{fig:membrane_linking_projection}. The directions $t_{1,2,3}$ show the time ordering of contributions to projection from worldsheets $G,H,F$, so clarify to which $Tlk_{IJK}$ some triple point contributes (see after Eq.~\eqref{eq:tlk}). For example, at point $a$, direction of $t_1$, $t_2$, $t_3$ shows that worldsheet projection at this point comes from: $G$ rather than $G^{-1}$, $H^{-1}$ rather than $H$, $F^{-1}$ rather than $F$, respectively. Therefore point $a$ contributes to $Tlk_{FHG}$.}
  \label{fig:membrane_operator_commutation_projection}
\end{figure}

Returning to the 2+1D case, as briefly reviewed in the Introduction, there is another way to relate an $\mathcal{S}$ matrix element to a topological property of the quasiparticle braiding process. Namely, the quasiparticle tunneling operators obey a non-trivial algebra, and a certain product of these operators gives the identity operator times a complex number which equals an $\mathcal{S}$ matrix element.\cite{Wen:1990p5870,Oshikawa:2006p5869} Fig.~\ref{fig:2dbraid}c depicts such a product of tunneling operators as a time sequence of events where particle---antiparticle pairs tunnel across the periodic system, in the same fashion as in Fig.~\ref{fig:2dbraid}a. The four worldlines in this process can be connected to reveal two linked worldlines, Fig. ~\ref{fig:2dbraid}d. Notice that this procedure can be done in the representation of the system as a parallelepiped with periodic boundary conditions, without need for embedding the 2+1D space time in a three dimensional space. One again confirms that $\mathcal{S}$ matrix element describes braiding, which can in turn be seen as linking of worldlines.

Here we present an analogous interpretation of a 3+1D $\mathcal{S}$ matrix element.
%In this section, we will construct braiding process with the same triple linking number by membrane operators. After straightforward calculation, we will get the same result as $\mathcal{S}$ matrix element.
The algebra of membrane operators follows from their definition:
\begin{align}
  \label{eq:membrane_algebra} 
  F_{u,\lambda}^{(z)}H_{w,\mu}^{(x)}|a,v,c\rangle&=\frac{\tilde{\chi}_{\lambda}^{u,v}(wc)\tilde{\chi}_{\mu}^{v,w}(a)}{\gamma_{v,wc}(u,a)\gamma_{a,v}(w,c)}|ua,v,wc\rangle\\\notag
  H_{w,\mu}^{(x)}F_{u,\lambda}^{(z)}|a,v,c\rangle&=\frac{\tilde{\chi}_{\lambda}^{u,v}(c)\tilde{\chi}_{\mu}^{v,w}(ua)}{\gamma_{v,c}(u,a)\gamma_{ua,v}(w,c)}|ua,v,wc\rangle.\\\notag
  %G_{v,\mu}^{(x)}F_{u,\nu}^{(y)}|a,b,c\rangle&=\frac{\tilde{\chi}_{\nu}^{c,u}(b)\tilde{\chi}_{\mu}^{v,c}(ua)}{\gamma_{b,c}(u,a)\gamma_{c,ua}(v,b)}|ua,vb,c\rangle,\\\notag
  %F_{u,\nu}^{(y)}G_{v,\mu}^{(x)}|a,b,c\rangle&=\frac{\tilde{\chi}_{\mu}^{v,c}(a)\tilde{\chi}_{\nu}^{c,u}(vb)}{\gamma_{c,a}(v,b)\gamma_{vb,c}(u,a)}|ua,vb,c\rangle.
\end{align}
Consider the membrane operator product $\langle G^{-1}F^{-1}H^{-1}FHG\rangle$, where the expectation value is obtained in state $|a,e,c\rangle$ with $a,c$ arbitrary (note that such a state is in the ground state manifold, see after Eq.~\eqref{gamma_proj_rep}). Here we label $F=F_{u,\lambda}^{(z)}$, $G=\sum_{c,a}G_{c,a}^v$ and $H=H_{w,\mu}^{(x)}$ for simplicity. Using Eq.(\ref{eq:membrane_algebra}), it is straightforward to show
\begin{align}
  \label{eq:Smat_fromMemb}
  \langle G^{-1}F^{-1}H^{-1}FHG\rangle=\frac{\tilde{\chi}_{\lambda}^{u,v}(w)}{\tilde{\chi}_{\mu}^{v,w}(u)},
  %\langle H^{-1}G^{-1}F^{-1}GFH\rangle=\frac{\tilde{\chi}_{\mu}^{v,w}(u)}{\tilde{\chi}_{\nu}^{w,u}(v)}.
\end{align}
so the quantum amplitude equals $\mathcal{S}$ matrix element $\langle v,w,\mu|S|u,v,\lambda\rangle$ up to factor $|G|$!

As in the previous subsection, the membrane operators Eq.~\eqref{eq:membranesFG},~\eqref{eq:membranesHH},~\eqref{eq:membranesFGmapMES}, are interpreted as events of tunneling a flux-loop---antiflux-loop pair across a plane in the periodic system; for example, from Eq. ~\eqref{eq:membranesFG}, the $F_{u,\lambda}^{(z)}$ describes the tunneling of loop in $zy$ plane. The quantum amplitude in Eq.~\eqref{eq:Smat_fromMemb} can then be seen as the ``movie'' in Fig.~\ref{fig:membrane_operator_commutation_movie}.

Analogously to the 2+1D case, where two worldlines, defined by a particle tunneling operator and its inverse, were connected into a single worldline, we can consider that the events defined by $F$ and $F^{-1}$ form a single worldsheet, and so on for $G$ and $H$. Having exactly three worldsheets, we calculate their TLN. Projecting the ``movie'' onto the three dimensional space slice at time $t=-\infty$, we find eight triple intersection points of the projected worldsheets, Fig. ~\ref{fig:membrane_operator_commutation_projection}. For simplicity of presentation, we offset the spatial position of inserted operator and its inverse, i.e., the membrane is moved slightly between the time of its appearance and disappearance. We checked that this offset does not influence the result. Notice that the orientation of membrane operator and its inverse are opposite, see Appendix~\ref{app:geom}. A straightforward calculation from each triple point gives: $a,c,e:Tlk_{FHG}=1$, $b:Tlk_{HFG}=1$, $f: Tlk_{GHF}=-1$, $e,g,h: Tlk_{GFH}=-1$, which is exactly the same triple linking number obtained from $\mathcal{S}$ matrix calculation!

%{\color{red} OLD PART!}
% A key question now becomes: What is a robust physical characterization of the process captured by the non-trivial membrane operator algebra? The answer is that in this process worldsheets of loops, which are represented by membrane operators, have a non-trivial TLN. First, we denote the worldsheet components $H,F,G$ as $1,2,3$, respectively.
%To calculate the value of TLN, we project the 4D ``movie'' onto the three-dimensional slice at time $t=-\infty$, and find eight triple intersection points of the projected worldsheets, Fig.~\ref{fig:membrane_operator_commutation_projection}. For simplicity of presentation, we offset the spatial position of inserted operator and its inverse, i.e., the membrane is moved slightly between the time of its appearance and disappearance. We checked that this does not influence the result.
%A straightforward calculation from each triple point gives: $b:Tlk_{123}=1$, $a,e,f: Tlk_{132}=1$, $d: Tlk_{231}=-1$, $c,g,h: Tlk_{321}=-1$. The obtained values of $Tlk_{ijk}$ are consistent (Eq.~\eqref{eq:tlk}).
%The membrane expression is therefore characterized by $a=1, b=1$, see Eq.~\eqref{eq:1}.

\subsubsection{Braiding process of three flux loops}

\begin{figure}
  \includegraphics[width=0.25\textwidth]{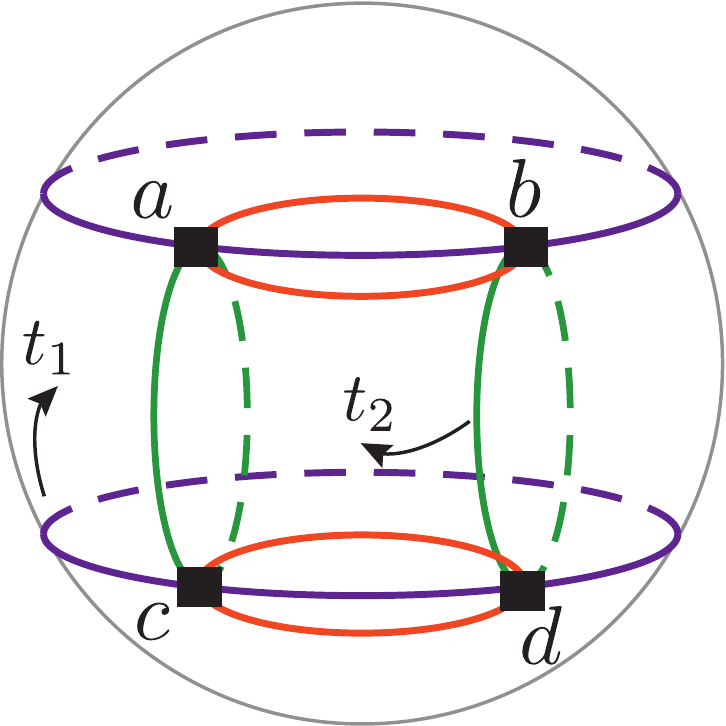}
  \caption{(color online) Movie in Fig. \ref{fig:membrane_linking_movie} projected to three-dimensional space at $t=-\infty$. Triple points are marked $a,b,c,d$. Lines show the pairwise intersections of projected worldsheets. Purple lines: for $F$ and $G$ worldsheets; Orange: for $F$ and $H$; Green: for $G$ and $H$. The projected $G$ worldsheet in this figure takes the form of the sphere; $H$ and $F$ take the form of tori (not shown). The directions $t_{1,2}$ show the time ordering of contributions to projection from worldsheets $G$, $H$, so clarify to which $Tlk_{IJK}$ some triple point contributes (see after Eq.~\eqref{eq:tlk}). For example, at point $c$, $t_2$ shows that projection of $H$ at this point comes at later times, so after $F$, while $t_1$ shows that $G$ comes from earlier times, so before both $F$ and $H$, altogether contributing to $Tlk_{HFG}$. This process has same triple linking number as the one in Fig.~\ref{fig:membrane_operator_commutation_projection}.}
  \label{fig:membrane_linking_projection}
\end{figure}

In previous subsections, we have interpreted the membrane operators as representing an instantaneous event of creating a loop---anti-loop pair and expanding the loop across a plane in the periodic system until the loops annihilate. However, this kind of worldsheet evolution can be smoothly deformed to represent a more physically clear process. We therefore make a movie of three-flux-loop braiding process that gives exactly the same nontrivial triple linking number as the membrane process, as shown in Fig.~\ref{fig:membrane_linking_movie}. By projecting this braiding movie, we get Fig.~\ref{fig:membrane_linking_projection}, in which it is straightforward to measure the TLN:

Triple point $a$ gives $\mathrm{Tlk}_{GHF}=-1$, triple point $b$ gives $\mathrm{Tlk}_{GFH}=-1$, triple point $c$ gives $\mathrm{Tlk}_{HFG}=1$ and triple point $d$ gives $\mathrm{Tlk}_{FHG}=1$. The obtained values of $Tlk_{IJK}$ for the three-flux-loop braiding exactly match the membrane calculation result.

\section{Examples}

Here we will present the example of $G=Z_2\times Z_2$ cohomological gauge theories. Since $H^4(G,U(1))=Z_2\times Z_2$, they can represent different topological orders. This will show how the loop statistics can distinguish different topological orders.

It is convenient to label group $G$ elements $a$ as $(a_1,a_2)$, where $a_i\in\{0,1\}$. Group multiplication rule $a\cdot b$ is defined as $(\langle a_1+b_1\rangle,\langle a_2+b_2\rangle)$, where we introduce notation $\langle x\rangle\equiv x \,\mathrm{mod}\, 2$.

Since the cohomology group is $H^4(Z_2\times Z_2,U(1))\cong Z_2\times Z_2$, it can be parametrized by 4-cocycles
\begin{align}
  \{\omega_{ij}|i,j=0,1\},
\end{align}
with multiplication rule
\begin{align}
  \omega_{ij}(a,b,c,d)\cdot\omega_{i'j'}(a,b,c,d)=\omega_{\langle i+i'\rangle\langle j+j'\rangle}(a,b,c,d).
\end{align}

The explicit form of these 4-cocycles is\cite{Chen:2014p7995}
\begin{align}
  \label{eq:Z2Z2_omega_def}
  \omega_{00}(a,b,c,d)&=1,\\\notag
  \omega_{01}(a,b,c,d)&=\exp[\frac{i\pi}{2}a_1b_2(c_2+d_2-\langle c_2+d_2\rangle)],\\\notag
  \omega_{10}(a,b,c,d)&=\exp[\frac{i\pi}{2}a_2b_1(c_1+d_1-\langle c_1+d_1\rangle)],\\\notag
  \omega_{11}(a,b,c,d)&=\omega_{01}(a,b,c,d)\cdot\omega_{10}(a,b,c,d).
\end{align}
It is straightforward to check that these $\omega$ indeed satisfy the 4-cocycle condition.

One can now work out the induced 3-cocycle $\beta_a$ and 2-cocycle $\gamma_{a,b}$ using their definitions in Eq.(\ref{eq:beta_def}) and Eq.(\ref{eq:gamma_def}). For induced 3-cocycle, we get
\begin{align}
  \label{eq:Z2Z2_beta_def}
  \beta_{00,a}(b,c,d)&=1,\\\notag
  \beta_{01,a}(b,c,d)&=\exp[\frac{i\pi}{2}(a_1b_2-a_2b_1)(c_2+d_2-\langle c_2+d_2\rangle)],\\\notag
  \beta_{10,a}(b,c,d)&=\exp[\frac{i\pi}{2}(a_2b_1-a_1b_2)(c_1+d_1-\langle c_1+d_1\rangle)],\\\notag
  \beta_{11,a}(b,c,d)&=\beta_{01,a}(b,c,d)\beta_{10,a}(b,c,d).
\end{align}
It follows that the 3-cocycle $\beta_a$ can be expressed as
\begin{align}
  \beta_a(b,c,d)=\exp\left[\frac{i\pi}{2} P_{ij}^{a}b_i(c_j+d_j-\langle c_j+d_j\rangle)\right],
\end{align}
where $P_{ij}^{a}$ is some integer matrix. According to Ref.\onlinecite{Propitius:1995p6856}, then the induced 2-cocycle must be a coboundary $\gamma_{a,b}(c,d)=\delta\,\varepsilon_{a,b}(c,d)$, where
\begin{align}
  \varepsilon_{a,b}(c)=\exp\left(\frac{i\pi}{2} P_{ij}^{a}b_ic_j\right).
\end{align}
Altogether, for inequivalent 4-cocycles we get the induced 2-cocycle as
\begin{align}
  \label{eq:Z2Z2_epsilon_def}
  \varepsilon^{00}_{a,b}(c)&=1,\\\notag
  \varepsilon^{01}_{a,b}(c)&=\exp[\frac{i\pi}{2}(a_1b_2c_2-a_2b_1c_2)],\\\notag
  \varepsilon^{10}_{a,b}(c)&=\exp[\frac{i\pi}{2}(a_2b_1c_1-a_1b_2c_1)],\\\notag
  \varepsilon^{11}_{a,b}(c)&=\varepsilon^{01}_{a,b}(c)\cdot\varepsilon^{10}_{a,b}(c).
\end{align}

Now, we are ready to calculate statistics of loops and particles. We will focus on 
\begin{align}
  \label{eq:SmatZ2Z2}
  &|G|\cdot\langle w,u,\nu|S|v,w,\mu\rangle=\frac{\tilde{\chi}_{\mu}^{v,w}(u)}{\tilde{\chi}_{\nu}^{w,u}(v)}\\\notag
  =&\frac{\chi_{\mu}(u)}{\chi_{\nu}(v)}\cdot\frac{\varepsilon_{v,w}(u)}{\varepsilon_{w,u}(v)}.
\end{align}
In the second equality, we have defined 
\begin{align}
  \tilde{\chi}_{\mu}^{vw}(u)&=\varepsilon_{v,w}(u)\cdot\chi_{\mu}(u),\\\notag
  \tilde{\chi}_{\nu}^{wu}(v)&=\varepsilon_{w,u}(v)\cdot\chi_{\nu}(v).
\end{align}
where $\chi_{\mu}(\chi_{\nu})$ is one-dimensional linear representation of $Z_2\times Z_2$. One can easily check the above definition of $\tilde{\chi}_{\mu}$ and $\tilde{\chi}_{\nu}$ is consistent, due to $\gamma_{a,b}$ being a 2-coboundary. Labeling $\mu=(\mu_1,\mu_2)$ as $Z_2\times Z_2$ group element,
\begin{align}
  \chi_{\mu}(u)=e^{i\pi(\mu_1u_1+\mu_2u_2)}=e^{i\pi\vec{\mu}\cdot\vec{u}}.
\end{align}

First, let us consider the case $w=(0,0)$. In this case, only the $\chi_\lambda$ factors are non-trivial in second line of Eq.~\eqref{eq:SmatZ2Z2}, which is interpreted as contribution from Aharonov-Bohm phase of braiding particles around flux-loops. In this case, the phase factor equals $e^{i\pi(\vec{\mu}\cdot\vec{u}-\vec{\nu}\cdot\vec{v})}$, which is independent of choice of cocycle. Namely, statistics between particles and loops cannot distinguish different phases.

Then, we turn to the general case. We get an additional phase factor $s_l$ beyond $e^{i\pi(\vec{\mu}\cdot\vec{u}-\vec{\nu}\cdot\vec{v})}$, and the $s_l$ factor comes from $\varepsilon$ in Eq.~\eqref{eq:SmatZ2Z2}. In other words, it is present even when $\mu=\nu=0$, i.e., $\chi$ representations are trivial, so there are no charged particles. Therefore, $s_l$ represents statistics of flux-loops. We list $s_l$ obtained from different 4-cocycles as follows
\begin{itemize}
  \item $\omega_{00}$: $s_l=1$.
  \item $\omega_{01}$: $s_l=e^{\frac{i\pi}{2}[(u_1v_2+u_2v_1)w_2-2u_2v_2w_1]}$.
  \item $\omega_{10}$: $s_l=e^{\frac{i\pi}{2}[(u_1v_2+u_2v_1)w_1-2u_1v_1w_2]}$.
  \item $\omega_{11}$: $s_l=e^{\frac{i\pi}{2}[(u_1v_2+u_2v_1)(w_1+w_2)-2u_1v_1w_2-2u_2v_2w_1]}$.
  \end{itemize}
We can see that flux-loop braiding can indeed distinguish different topological orders in 3+1D, recalling here that the membrane operator expression is identified with a particular type of three-flux-loop braiding. In particular, according to previous section we can identify the flux-loops (blue, red, black) in Fig.~\ref{fig:membrane_linking_movie} with fluxes $(u,v,w)$ here, and there are no charges present.

Now, we turn to $\mathcal{T}$ matrix element $\tilde{\chi}_{\lambda}^{u,v}(u)=\langle u,v,\lambda|T^{31}|u,v,\lambda\rangle$. In the same way as above, we get
\begin{itemize}
  \item $\omega_{00}$: $\tilde{\chi}_{\lambda}^{u,v}(u)=e^{i\pi\vec{\lambda}\cdot\vec{u}}$.
  \item $\omega_{01}$: $\tilde{\chi}_{\lambda}^{u,v}(u)=e^{i\pi\vec{\lambda}\cdot\vec{u}}e^{\frac{i\pi}{2}(u_1v_2u_2-u_2v_1u_2)}$.
  \item $\omega_{10}$: $\tilde{\chi}_{\lambda}^{u,v}(u)=e^{i\pi\vec{\lambda}\cdot\vec{u}}e^{\frac{i\pi}{2}(u_2v_1u_1-u_1v_2u_1)}$.
  \item $\omega_{11}$: $\tilde{\chi}_{\lambda}^{u,v}(u)=e^{i\pi\vec{\lambda}\cdot\vec{u}}e^{\frac{i\pi}{2}(u_1v_2+u_2v_1)(u_1-u_2)}$.
\end{itemize}
While the $e^{i\pi\vec{\lambda}\cdot\vec{u}}$ can be interpreted as Aharonov-Bohm phase of particles going around loop, the remaining part also encodes information about loop statistics. While we do not have a proof at this time, we believe that this phase is related to the ribbon nature of flux loop, or in other words to a thickness of the membrane.

\section{Discussion and conclusions}

One of our main results is the construction of MES states on the three-torus for the 3+1D cohomological gauge theory, which can be trivially generalized to arbitrary number of unit-cells. The $S,T$ transformation matrices take a simple form in this basis.

We discussed that the $S$-matrix elements are directly related to the braiding of loop excitations. The $T$-matrix elements, which are diagonal in the MES basis, correspond to the generalization of topological spin for loop excitations. Here physically the loop excitations are generally expected to be ribbon excitations with two different loop-edges. We expect that the geometrical interpretation of the $T$-matrix elements is related to the braiding involving different loop-edges.

Although we use exactly solvable models and 3+1D topological quantum field theories to compute their $S,T$ matrices, these 3+1D $S,T$ matrices are in principle measurable quantities in practical model Hamiltonians. In particular, given a topologically ordered phase in 3+1D with its topologically degenerate ground sector on three-torus $T^3$, one can firstly find a MES basis, similarly to the algorithms proposed in 2+1D.\cite{Zhang:2012p7534} For instance, for the $S$-matrix element between two MES $|\Xi_i\rangle$ and $|\Xi_j\rangle$: $S_{ij}$, one can perform the following thought numerical measurement. Because the topological properties do not depend of local geometry, we can assume that these ground states live on a cube with periodic boundary conditions. Then one can consider the state rotated by 120$^\circ$ along the (111) direction of the cube: $R_{120^\circ}|\Xi_i\rangle$. Because $R_{120^\circ}|\Xi_i\rangle$ and $|\Xi_j\rangle$ belong to the same topological phase, in the absence of symmetry there should exist a Hamiltonian path $H(\tau)$ ($\tau\in[0,1]$) such that $|\Xi_j\rangle$($|\Xi_j\rangle$) are the ground state of $H(0)$($H(1)$), and the ground state sectors of $H(\tau)$ are adiabatically connected. One can then define a projection operator $\hat P_{\tau}$ into the ground state sector of $H(\tau)$ for any given $\tau$. The many-body quantum amplitude related to the adiabatic time-evolution process of the $S$-transformation can be computed as $\langle\Xi_j|\hat P_{N-1/N}\cdot...\cdot\hat P_{2/N}\cdot\hat P_{1/N}R_{120^\circ}|\Xi_i\rangle$ as $N\rightarrow \infty$. This computation is a realization of the topological quantum field theory time-evolution. 

We expect that this quantum amplitude is related to the $S$-matrix elements $s_{ij}$ at most by an overall ambiguity $U(1)$ phase $e^{i\theta}$, which is due to the nonuniversal local physics in the time-evolution, and a phase $e^{i\phi_i-i\phi_j}$ which is due to the gauge choice of $|\Xi_i\rangle$,$|\Xi_j\rangle$. Even with these ambiguities, such measurements can still be used to extract useful information about the $S,T$ matrices which potentially could fully determine them.

Recently, there has been a lot of progress in relating topologically ordered phases to symmetry protected topological (SPT) and symmetry enriched topological (SET) phases, for example by partially or completely ungauging the gauge group $G$, i.e., by transformations between global and local symmetries.\cite{Levin:2012p7190,Mesaros:2013p7698,Chen:2014p8001,Gu:2014p8002,Ye:2013p8003,Chen:2014p8000,Hung:2012p7997,Zaletel:2013p7999,Xu:2013p7998} We therefore expect that our work will be useful in characterization of SPT and SET phases too.

Our discussion has been limited to the case in which both group $G$ and its projective representations are Abelian. These constraints are introduced here for simplicity rather than due to difficulty of principle. In the most general case, we expect $\tilde{\chi}$ to be the character of a projective representation, while modular transformations can easily be generalized, as shown in Ref.\onlinecite{Hu:2012p7528} for 2+1D case. In those cases, three-loop braiding may give a unitary matrix which corresponds to non-Abelian statistics of loops. In fact, we expect that the triple linking is relevant for any 3+1D topological order, beyond cohomological models. Namely, as long as the MESs in 3+1D topologically ordered state can be written using tunneling operators of loop-like excitations, which we believe should always be the case, our general geometrical approach based on space-time embedding shows that MES overlaps, which appear as S-matrix elements, will involve the triple linking of worldsheets defined by those tunneling operators.

Finally, let us consider a trivial but ubiquitous example of $G=Z_2$. In this case, $H^4(G,U(1))=Z_1$, so the cocycle can be set to identity map. The braiding phase $\tilde{\chi}_{\mu}^{vw}(u)/\tilde{\chi}_{\nu}^{wu}(v)$ reduces to a linear representation $\chi_{\mu}(u)/\chi_\nu(v)$, where group elements $u,v=0,1$, and $\mu,\nu=0,1$ label the representations of $Z_2$:
\begin{align}
  \chi_{\mu}(u)=e^{i\pi \mu u}.
\end{align}
The braiding phase therefore equals $e^{i\pi(\mu u-\nu v)}$. There is no contribution from flux-loop braiding, since the 1-cocycle factors in Eq.~\eqref{eq:S_act_MESfact} are trivial. In summary, the modular $S$ transformation for common $Z_2$ gauge theory in 3+1D tells us that particles see a flux-loop as a $\pi$-flux, and the flux-loops themselves have trivial braiding.

Using the MES basis and Eq.~\eqref{eq:T_act_MES},~\eqref{eq:S_act_MES}, we directly obtain the $\mathcal{S}$ and $\mathcal{T}$ matrices of 3+1D $Z_2$ theory in their canonical form:
\begin{align}\notag
  \mathcal{S}&=\frac{1}{2}
  \begin{pmatrix}
 1 & 1 & 0 & 0 & 1 & 1 & 0 & 0 \\
 1 & 1 & 0 & 0 & -1 & -1 & 0 & 0 \\
 1 & -1 & 0 & 0 & 1 & -1 & 0 & 0 \\
 1 & -1 & 0 & 0 & -1 & 1 & 0 & 0 \\
 0 & 0 & 1 & 1 & 0 & 0 & 1 & 1 \\
 0 & 0 & 1 & 1 & 0 & 0 & -1 & -1 \\
 0 & 0 & 1 & -1 & 0 & 0 & 1 & -1 \\
 0 & 0 & 1 & -1 & 0 & 0 & -1 & 1 \\
  \end{pmatrix}\\\notag
\mathcal{T}^{31}&=\mathrm{Diag}(1,1,1,1,1,-1,1,-1),
\end{align}
where the MES basis $|a,b,\lambda\rangle$, with $a,b,\lambda\in\{0,1\}$, is here naturally ordered according to binary numbers with digits $ab\lambda$. These matrices are consistent with the $\mathcal{S}$ and $\mathcal{T}$ matrices derived for the same theory in Ref.\onlinecite{Moradi:2014p7978}.

During the preparation of this manuscript we noticed a very recent paper that also considers aspects of flux-loop braiding in 3+1D.\cite{Wang:2014p7974}

This work was supported by the Alfred P. Sloan foundation and National Science Foundation under Grant No. DMR-1151440.

\appendix

\section{Definition of the cohomology group and projective representations}
\label{app:coh}
We begin with a brief introduction to group cohomology. In this paper, we will not present the most general definition of group cohomology.

For a finite group $G$, and an abelian group $M$ ($M$ does not need to be finite or discrete), one can consider an arbitrary function that maps n elements of $G$ to an element in $M$; $\omega: G^n\rightarrow M$ or equivalently $\omega(g_1,g_2,...,g_n)\in M$, $\forall g_1,g_2,...g_n\in G$. Such a group function is called an n-cochain. The set of all n-cochains, which is denoted as $C^n(G,M)$, forms an abelian group in the usual sense: $(\omega_1\cdot\omega_2)(g_1,g_2,...,g_n)=\omega_1(g_1,g_2,...,g_n)\cdot \omega_2(g_1,g_2,...,g_n)$, in which the identity n-cochain is a group function whose value is always the identity in $M$.

One can define a mapping $\delta$ from $C^n(G,M)$ to $C^{n+1}(G,M)$: $\forall \omega\in C^n(G,M)$, define $\delta \omega \in C^{n+1}(G,M)$ as
\begin{align}\label{eq:n-cocycle}
\delta &\omega (g_1,...,g_{n+1})=\omega(g_2,...,g_{n+1})\cdot\omega^{(-1)^{n+1}}(g_1,...,g_n)\notag\\
&\times \prod_{i=1}^n \omega^{(-1)^i}(g_1,..,g_{i-1},g_i\cdot g_{i+1},g_{i+2},..,g_{n+1}).
\end{align}
It is easy to show that the mapping $\delta$ is nilpotent: $\delta^2 \omega=1$ (here $1$ denotes the identity (n+2)-cochain). In addition, for two n-cochains $\omega_1,\omega_2$, obviously $\delta$ satisfies $\delta(\omega_1\cdot\omega_2)=(\delta\omega_1)\cdot(\delta\omega_2)$.

An n-cochain $\omega(g_1,...g_n)$ is called an n-cocyle if and only if it satisfies the condition: $\delta\omega =1$, where $1$ is the identity element in $C^{n+1}(G,M)$. When this condition is satisfied, we also say that $\omega(g_1,...g_n)$ is an n-cocycle of group $G$ with coefficients in $M$. The set of all n-cocycles, denoted by $Z^n(G,M)$, forms a subgroup of $C^{n}(G,M)$.

Not all different cocyles are inequivalent. Below we define an equivalence relation in $Z^n(G,M)$. Because $\delta$ is nilpotent, for any (n-1)-cochain $c(g_1,...,g_{n-1})$, we can find the n-cocyle $\delta c$. And if an n-cocyle $b$ can be represented as $b=\delta c$, for some $c\in C^{n-1}(G,M)$, $b$ is called an n-coboundary. The set of all n-coboundaries, denoted by $B^n(G,M)$, forms a subgroup of $Z^n(G,M)$. Two n-cocycles $\omega_1,\omega_2$ are equivalent (denoted by $\omega_1\sim\omega_2$) if and only if they differ by an n-coboundary: $\omega_1=\omega_2\cdot b$, where $b\in B^n(G,M)$.

The n-th cohomology group of group $G$ with coefficients in $M$, $H^{n}(G,M)$, is formed by the equivalence classes in $Z^n(B,M)$. More precisely: $H^{n}(G,M)=Z^n(G,M)/B^n(G,M)$.

In this paper we make a lot of use of 4-cocycles $\omega$. We always choose them to be in ``canonical'' form, which means that $\omega(g_1,g_2,g_3,g_4)=1$ if any of $g_1,g_2,g_3,g_4$ is equal to $\openone$ (the identity element of group $G$). For any of the inequivalent cocycles mentioned above, it is always possible to choose a gauge such that $\omega$ becomes canonical~\cite{Chen:2013p6670}.

In usual unitary group representations, each group element $g$ in $G$ is represented by a unitary matrix $D(g)$, which satisfies: $D(g_1)\cdot D(g_2)=D(g_1\cdot g_2)$. The projective representations of the group $G$ are defined by modifying this relation by a phase factor   $\omega(g_1,g_2)\in U(1)$:
\begin{equation}
  \label{eq:6}
D(g_1)\cdot D(g_2)=\omega(g_1,g_2)D(g_1\cdot g_2),
\end{equation}
where $\omega(g_1,g_2)$ is a function of $g_1,g_2$ called a factor system.\cite{Isaacs1976} A factor system cannot be arbitrary. In order to satisfy the associativity condition: $[D(g_1)\cdot D(g_2)]\cdot D(g_3)=D(g_1)\cdot [D(g_2)\cdot D(g_3)]$, the factor system must satisfy the equation
\begin{align}
 \omega(g_1,g_2)\cdot \omega(g_1\cdot g_2,g_3)=\omega(g_2,g_3)\cdot \omega(g_1,g_2\cdot g_3).\label{eq:2-cocycle}
\end{align}
This relation is precisely the condition for $\omega$ to be a 2-cocycle (the condition is $\delta\omega=1$ in Eq.(\ref{eq:n-cocycle}) for $n=2$). If $\omega(g_1,g_2)$ is a 2-coboundary, it can be written as $\omega(g_1,g_2)=c(g_1)\cdot c(g_2)/c(g_1\cdot g_2)$ for a certain 1-cochain $c(g)$. If two 2-cocyles, $\omega_1,\omega_2$ , differ by a 2-coboundary:
\begin{equation}
  \label{eq:gauge2coc}
  \omega_1(g_1,g_2)=\omega_2(g_1,g_2)\cdot\frac{ c(g_1)\cdot c(g_2)}{c(g_1\cdot g_2)},
\end{equation}
it is obvious that they correspond to equivalent projective representations, because one can absorb the 1-cochain into $D(g)$ by redefining $\tilde D(g)=c(g)\cdot D(g)$, after which the two factor systems becomes the same (this is actually the definition of equivalent projective representations). Therefore the $H^2(G,U(1))$ also classifies all inequivalent (factor systems of) projective representations.

\section{Geometrical interpretation of $\mathcal{S}$ matrix and triple linking calculation conventions}
\label{app:geom}

In this section we present the details of embedding the 3+1D space-time in a four-dimensional Euclidean space $\mathbb{R}^4$, which is applied to the third line of Eq.~\eqref{eq:S_matrix_MES_z} to show that in the embedded space worldsheets exhibit triple linking.
% it to the case when To see the meaning of $\mathcal{S}$ matrix elements, let us try to interpret the above equation as a spacetime event.
The embedding is a generalization of the 2+1D case shown in Fig.~\ref{fig:2dbraid}a,\ref{fig:2dbraid}b. The space manifold is a three-torus $T^3$ topologically a product of three circles $S_{(1)}^1\times S_{(2)}^1\times S_{(3)}^1$ with $1,2,3$ the spatial directions. At a given moment of time this space is embedded in $\mathbb{R}^4$ to form the manifold $\Sigma$ which is simply the surface of a three-torus:
\begin{align}
  \label{eq:3-torus}
  X=&[r_1+(r_2+r_3 \cos\gamma)\cos\beta]\cos\alpha,\\\notag
  Y=&[r_1+(r_2+r_3\cos\gamma)\cos\beta]\sin\alpha,\\\notag
  Z=&(r_2+r_3\cos\gamma)\sin\beta,\\\notag
  W=&r_3\sin\gamma,
\end{align} 
where $r_i$ is radius for $S_{(i)}^1$, which is constant at fixed time, and $(X,Y,Z,W)$ are Cartesian coordinates in $\mathbb{R}^4$. The $\alpha,\beta,\gamma$ are angles of $S_{(1)}^1,S_{(2)}^1,S_{(3)}^1$, respectively, and correspond to Cartesian spatial coordinates in a cube with periodic boundary conditions. Without loss of generality, for every moment in time $t$ we set $r_1>r_2>r_3$. The embedding of different time slices is chosen such that the manifold $\Sigma$ at later times always contains all $\Sigma$ from the past. More precisely, $[\Sigma\times(-\infty,t_1)]\subset[\Sigma\times(-\infty,t_2)]$, for any $t_1<t_2$. Finally, the four-dimensional manifold $M=\Sigma\times(-\infty,\infty)$ covers the whole $\mathbb{R}^4$ space.

As $t\to-\infty$, $r_2,r_3\to0$, so the space manifold shrinks to a circle in $XY$ plane: $X^2+Y^2=r_1[t=-\infty]^2$. As $t\to\infty$, $r_1,r_2,r_3\to\infty$, and the space manifold $\Sigma$ asymptotically approaches the $W$-axis, which can be viewed as the $S_{(3)}^1$ circle ($\gamma$ circle) with $r_3=\infty$. Using all time $(-\infty,\infty)$ with these asymptotic limits in the embedding is useful for consistently removing ambiguities in the embedding of events corresponding to membrane operators, as will soon become clear.

After embedding the space-time manifold itself, let us proceed to an $\mathcal{S}$ matrix element as a sequence of space-time events. For convenience, we repeat Eq.~\eqref{eq:S_matrix_MES_z} here:
\begin{align}
  \label{eq:4}
  \langle v,w,\mu|S|u,v,\lambda\rangle=&\langle\mu,v,w|u,v,\lambda\rangle=\\\notag
  =&\langle1,e,e|(G_{v,1}^{(x)})^{-1}(H_{w,\mu}^{(x)})^{-1}F_{u,\lambda}^{(z)}G_{v,1}^{(z)}|e,e,1\rangle.
\end{align}
It is useful to think about this equation as a two-step process in time: from $t=-\infty$ to $t=0$ and from $t=0$ to $t=\infty$. At $t=0^-$, the system is in state $|u,v,\lambda\rangle$, obtained by insertion of membranes $G_{v,1}^{(z)}$ and $F_{u,\lambda}^{(z)}$ at some time $t<0$ into the trivial MES $|e,e,1\rangle$. On the other hand, the interval from $t=0$ to $t=\infty$ can be interpreted as the conjugate of an appropriate history from $t=-\infty$ to $t=0$. This gives us the bra $\langle\mu,v,w|$ by simply inserting membranes $G$ and $H$ into the trivial MES at $t=\infty$. Joining the two histories at $t=0$, the product of bra and ket is obtained. So the quantum amplitude which equals the $\mathcal{S}$ matrix element is naturally interpreted as a sequence of space-time events.

Crucially, the coordinates $x,y,z$ used for the membranes are in the original three-dimensional space, and we yet have to consistently identify them with the $\alpha,\beta,\gamma$ coordinates which were defined above in the embedding of space into $\mathbb{R}^4$.
% It remains to precisely define how the events corresponding to insertion of membrane operators appear in the embedded space, which requires a consistent identification between the space coordinates $x,y,z$ and $\alpha,\beta,\gamma$ we introduced.
It turns out that $F$ creates a sheet covering $S_{(1)}^1\times S_{(2)}^1$, $G$ creates the sheet $S_{(3)}^1\times S_{(1)}^1$ and membrane $H$ covers $S_{(2)}^1\times S_{(3)}^1$, as we show below. In other words, if we cut open the three-torus $T^3$ to a cube, then $x$ axis corresponds to $\gamma$, $y$ axis corresponds to $\beta$ and $z$ axis corresponds to $\alpha$, remembering that $F$ is membrane in the $yz$ plane, $G$ is membrane in the $zx$ plane and $H$ is membrane in $xy$ plane.

To see this, we start from the trivial MES ket $|e,e,1\rangle=\frac{1}{\sqrt{|G|}}\sum_c|e,e,c\rangle$, where the three labels inside the ket correspond to the three directions $x,y,z$. Consider the corresponding limit $t\to-\infty$, in which the $\beta,\gamma$ circles shrink to points, while the $\alpha$ circle remains at finite radius $r_1$. Therefore the product of group elements along the $\alpha$ circle remains unconstrained in this limit. On the other hand, we conclude that the product of group elements along $\beta$ or $\gamma$ direction has to be identity $e$ to be shrinkable to a point. More concretely, consider a consistent triangulation of entire $\mathbb{R}^4$, so that the nested $\Sigma$ manifolds coming from different times are connected by the consistent triangulation. Then the $\beta$ and $\gamma$ circles at times $t<0$ are bounding consistently triangulated discs, and as they shrink the zero-flux rule through the discs would force the product of group elements along the circles to identity $e$. Altogether, only the $\alpha$ coordinate can be identified as $z$, while $\beta,\gamma$ correspond to $x,y$ (not necessarily in that order).
%So, we conclude that $z$ axis corresponds to circle $S_{(1)}^1$.

We can perform a similar analysis for bra $\langle1,e,e|$ and the corresponding limit $t\to\infty$. The finite loop at $t=\infty$ is $S_{(3)}^1$, the $\gamma$ circle. Then, it is easy to confirm that $x$ corresponds to $\gamma$. So, the remaining axis $y$ corresponds to $S_{(2)}^1$, i.e., to the $\beta$ circle.

Finally we discuss some conventions we have chosen in the calculations of TLN in both Section~\ref{sec:braid_S} and \ref{sec:braid_memb}. As described after Eq.~\eqref{eq:tlk}, the TLN requires the calculation of normal vectors to worldsheets which are projected from four to three dimensions. The overall sign of $Tlk_{IJK}$ therefore depends on precise definition of orientations of worldsheets, but these orientations are not inherent in the membrane operators and we need to choose them consistently. The $F$ membrane operator defines a sheet in the $yz$ plane, and we assign it the ordered pair $(z,y)$. After a projection to three dimensions, the derivatives with respect to $z,y$ define the sheet tangent vectors $\vec{n}_1,\vec{n}_2$, respectively, and the normal vector to sheet $F$ is chosen as $\vec{n}_1\times \vec{n}_2$. For $F^{-1}$, the normal is defined with opposite sign. In Section~\ref{sec:braid_memb}, the total worldsheet composed of $F$ and $F^{-1}$ therefore has a consistent normal vector throughout. In the exact same fashion, we assign to $G$ sheet the pair $(x,z)$, and to $H$ the $(y,x)$ pair. Finally, we note that the Jacobian of the transformation in Eqs.~\eqref{eq:3-torus} has a negative determinant, so the embedding reverses the handedness of the coordinates in the three-dimensional projected space. Due to this, the TLN values calculated from the embedding are additionally multiplied by $-1$.

\bibliography{3d_modular}

\end{document}